\documentclass[journal]{IEEEtran}

\ifCLASSINFOpdf

\else

\fi

\usepackage{mathrsfs}
\usepackage{amsmath}
\usepackage{amsfonts}
\usepackage{amsthm}
\usepackage{amssymb}
\usepackage{graphicx}
\usepackage{subfigure}
\usepackage{indentfirst}
\usepackage{array}
\usepackage{epstopdf}
\usepackage{cite}
\usepackage{enumerate}
\usepackage{bm}
\usepackage[linesnumbered,ruled,vlined]{algorithm2e}
\usepackage{setspace}
\usepackage{multirow}
\usepackage{verbatim}
\usepackage{stfloats}
\usepackage{color}
\usepackage{makecell}

\SetKwComment{Comment}{//}{}
\SetKwBlock{Begin}{Function}{end}

\newtheorem{proposition}{\bf{Proposition}}

\begin{document}

\title{Learning to Decode Protograph LDPC Codes}

\author{Jincheng~Dai,~\IEEEmembership{Member,~IEEE},
        Kailin~Tan,~\IEEEmembership{Student Member,~IEEE},
        Zhongwei~Si,~\IEEEmembership{Member,~IEEE},
        Kai~Niu,~\IEEEmembership{Member,~IEEE},
        Mingzhe~Chen,~\IEEEmembership{Member,~IEEE},
        H.~Vincent~Poor,~\IEEEmembership{Life Fellow,~IEEE},
        and Shuguang~Cui,~\IEEEmembership{Fellow,~IEEE}

\thanks{This work was supported in part by the National Natural Science Foundation of China under Grants 62001049 and 92067202, in part by the China Post-Doctoral Science Foundation under Grant 2019M660032, in part by Qualcomm Inc., in part by the U.S. National Science Foundation under Grant CCF-1908308, in part by the National Key R\&D Program of China under Grant 2018YFB1800800, in part by the Key Area R\&D Program of Guangdong Province under Grant 2018B030338001, in part by Shenzhen Outstanding Talents Training Fund, and in part by Guangdong Research Project under Grant 2017ZT07X152. \emph{(Corresponding author: Jincheng Dai; Kai Niu.)}}

\thanks{J. Dai, K. Tan, Z. Si, and K. Niu are with the Key Laboratory of Universal Wireless Communications, Ministry of Education, Beijing University of Posts and Telecommunications, Beijing, 100876 China (e-mail: daijincheng@bupt.edu.cn; tankailin@bupt.edu.cn; sizhongwei@bupt.edu.cn; niukai@bupt.edu.cn).}
\thanks{M. Chen is with the Department of Electrical Engineering, Princeton University, Princeton, NJ, 08544 USA, and also with the Shenzhen Research Institute of Big Data and Future Network of Intelligence Institute (FNii), the Chinese University of Hong Kong, Shenzhen, 518172 China (e-mail: mingzhec@princeton.edu).
}
\thanks{H. V. Poor is with the Department of Electrical Engineering, Princeton University, Princeton, NJ, 08544, USA (e-mail: poor@princeton.edu).}
\thanks{S. Cui is with the Shenzhen Research Institute of Big Data and Future Network of Intelligence Institute (FNii), the Chinese University of Hong Kong, Shenzhen, China, 518172 (e-mail: shuguangcui@cuhk.edu.cn).}
\thanks{Color versions of one or more figures in this article are available at xx}
\thanks{Digital Object Identifier xx}
\vspace{0em}
}

\maketitle

\begin{abstract}
The recent development of deep learning methods provides a new approach to optimize the belief propagation (BP) decoding of linear codes. However, the limitation of existing works is that the scale of neural networks increases rapidly with the codelength, thus they can only support short to moderate codelengths. From the point view of practicality, we propose a high-performance neural min-sum (MS) decoding method that makes full use of the lifting structure of protograph low-density parity-check (LDPC) codes. By this means, the size of the parameter array of each layer in the neural decoder only equals the number of edge-types for arbitrary codelengths. In particular, for protograph LDPC codes, the proposed neural MS decoder is constructed in a special way such that identical parameters are shared by a bundle of edges derived from the same edge-type. To reduce the complexity and overcome the vanishing gradient problem in training the proposed neural MS decoder, an iteration-by-iteration (i.e., layer-by-layer in neural networks) greedy training method is proposed. With this, the proposed neural MS decoder tends to be optimized with faster convergence, which is aligned with the early termination mechanism widely used in practice. To further enhance the generalization ability of the proposed neural MS decoder, a codelength/rate compatible training method is proposed, which randomly selects samples from a set of codes lifted from the same base code. As a theoretical performance evaluation tool, a trajectory-based extrinsic information transfer (T-EXIT) chart is developed for various decoders. Both T-EXIT and simulation results show that the optimized MS decoding can provide faster convergence and up to 1dB gain compared with the plain MS decoding and its variants with only slightly increased complexity. In addition, it can even outperform the sum-product algorithm for some short codes.
\end{abstract}


\IEEEpeerreviewmaketitle

\section{Introduction}\label{section_introduction}


\IEEEPARstart{R}{ecently}, low-density parity-check (LDPC) codes have been selected as the coding scheme for data channels in the 5G new radio (NR) system \cite{5G_NR_std_magazine,5G_NR_std}. LDPC codes utilize iterative decoding \cite{BP_1,BP_2} to achieve performance close to the Shannon limit \cite{LDPC_0045}. A number of iterative decoding algorithms exist. Among these decoding algorithms, the standard sum-product (SP) algorithm \cite{Lin_shu_channel_coding}, also called the belief-propagation (BP) algorithm \cite{J_Pearl_BP}, can achieve the optimal performance while its high decoding complexity hinders its practical use. Instead of the SP algorithm, the min-sum (MS) algorithm \cite{minsum} and its variants, such as the normalized min-sum (NMS) and the offset min-sum (OMS) \cite{minsum_variants,minsum_DE,minsum_adaptive}, have been developed to achieve the approximate the performance of the SP algorithm with much lower complexity, and thus they have been widely employed in practical systems. Multi-edge type LDPC (MET-LDPC) codes are introduced as a unified framework of structured LDPC codes in \cite{muti_edge_LDPC}. Protograph LDPC codes \cite{protograph_LDPC} define a subclass of MET-LDPC codes, which rely on the expansion of a smaller matrix or graph prototype (the base graph) into a full matrix or graph. With a well-designed structure, protograph LDPC codes can achieve better performance and are more suitable for an efficient encoding/decoding implementation.

In recent years, machine learning approaches have been rapidly developing \cite{deep_learning_chenmingzhe1,deep_learning_chenmingzhe2}, and they have given impressive performance in physical layer communications \cite{deep_learning_intro1,deep_learning_intro2,deep_learning_intro3}, e.g., channel estimation, decoding, etc. Particularly, in \cite{deep_learning_decoding_Allerton}, learnable weights are added to the SP algorithm and tuned by the gradient-based optimization method. The resulting weighted SP decoder in \cite{deep_learning_decoding_Allerton} can efficiently mitigate the negative impact caused by the short cycles in the Tanner graph, and shows the improvements of up to 1.5dB in the signal-to-noise ratio (SNR) against the conventional SP algorithm when decoding high-density parity-check (HDPC) codes. In \cite{deep_learning_decoding_ISIT} and \cite{deep_learning_decoding_JSTSP}, the NMS/OMS decoders with learnable normalizing/offset factors are studied to provide more hardware-friendly neural decoders. Nevertheless, all these existing works \cite{deep_learning_decoding_Allerton,deep_learning_decoding_ISIT,deep_learning_decoding_JSTSP} are on HDPC with codelengths less than 200 bits.

As the codelength increases, the dimensions of the parity-check matrix also increase, thus increasing the parameter array of the neural decoder, and resulting in extremely high training complexity. In addition, the iterative decoding convergence rate for longer codes becomes slower, which requires more iterations (e.g., 50, or even more) corresponding to quite deep neural networks. It may lead to the vanishing gradient problem during the training process. These reasons lead to that current neural SP decoders \cite{deep_learning_decoding_Allerton,deep_learning_decoding_ISIT,deep_learning_decoding_JSTSP} may only support short to moderate codelengths.

In this paper, from the point view of practicality, we propose a neural MS decoding method for protograph LDPC codes that makes full use of the lifting structure to overcome the limitation of codelength. We add fine-tuned parameters to the plain MS algorithm, making it a better approximation to the SP algorithm. The proposed neural MS decoder is constructed in a special way that follows a \emph{parameter-sharing mechanism}. By this means, only a small parameter array is required, whose size is only proportional to the number of edges in the base graph. This parameter-sharing mechanism efficiently reduces the training complexity and memory cost, thus overcoming the limitation of codelength in the conventional neural iterative decoders \cite{deep_learning_decoding_JSTSP}. In addition, different parameters are applied to different edge-types and iterations, which enables the neural MS decoder to mitigate the message correlation due to short cycles in the Tanner graph and tends to faster convergence, especially for some short codelength cases.

Note that the LDPC coding scheme in 5G NR \cite{5G_NR_std} adopts the protograph structure, including two base graphs (BG1 and BG2), to meet the requirements of high performance, high throughput, and low decoding latency. Hence, in this paper, we choose 5G LDPC codes as a representative of the protograph LDPC code class to train and verify the proposed neural MS decoder.

To the best of our knowledge, this is the first work to investigate neural decoding methods for protograph LDPC codes. The novelty and contribution of this paper are summarized as follows:

\begin{itemize}
  \item \emph{Parameter-Sharing Mechanism:} For protograph LDPC codes, since the base graph encapsulates the desired macroscopic structure, the original structural properties imposed by the base graph remain unchanged in the lifted graph. Because of this, the same parameters can be shared by a bundle of edges derived from the same edge in the base graph. Accordingly, the number of parameter pairs required for training only equals the number of edges in the base code. Moreover, the same parameter settings can be employed for decoding multiple codes derived from the same base code. This parameter-sharing mechanism makes it much more memory-efficient for the practical implementation of the proposed neural MS decoder.

  \item \emph{Iteration-by-Iteration Greedy Training:} Departing from the well-known multi-loss training method \cite{deep_learning_decoding_JSTSP}, we propose an iteration-by-iteration, i.e., layer-by-layer in neural networks, greedy training method by which only parameters of the last decoding iteration are learnable, and those for previous iterations are fixed. This training process is beneficial in three ways: (i) reducing the training complexity and combat the vanishing gradient problem in deep neural MS decoders; (ii) enabling the proposed neural MS decoder to mitigate the message correlation due to short cycles and tend to faster convergence, especially for some short codelength cases. This is aligned well with the early termination mechanism widely used in practical iterative decoders; and (iii) enabling the proposed neural MS decoder, with one iteration as its training granularity, to flexibly set up the decoding configurations, i.e., the number of iterations. In this way, many parameters in the optimized MS decoder can be reused rather than retrained for every different configuration.

  \item \emph{Codelength/Rate Compatible Training:} Taking advantage of the proposed parameter sharing mechanism, to enhance the generalization ability of a trained neural MS decoder to multiple codelengths and rates, we design a training process that randomly selects samples from a set of codes with different lengths or rates, and the codes provided for selection are derived from the same base code. This training process mitigates the overfitting problem to a certain code length (implicitly, the lifting way of protograph codes) or code rate, thus improving the length/rate compatibility of a trained neural decoder.

  \item \emph{T-EXIT Evaluation:} A trajectory-based extrinsic information transfer (T-EXIT) analysis is performed for neural MS decoders. The T-EXIT computation takes into account different normalizing and offset factors on edge bundles, which permits the convergence evaluation among various decoding algorithms. It presents the superiority of neural MS decoding from theoretical perspective.
\end{itemize}


The remainder of the paper is organized as follows. Section \ref{section_preliminaries} briefly reviews the mainstream iterative decoding algorithms and the protograph LDPC codes. Section \ref{section_neural_min_sum} describes the proposed neural MS decoding algorithm and its training method. In Section \ref{section_damping}, damping factors are further introduced to enhance the performance of neural MS decoders. Section \ref{section_T_EXIT} introduces the T-EXIT performance analysis method. Section \ref{section_performance_evaluation} shows the performance evaluation results with T-EXIT and simulations. Finally, Section \ref{section_conclusion} concludes the paper.

\section{Preliminaries}\label{section_preliminaries}

\subsection{Notational Conventions}

In this paper, we use calligraphic characters, such as ${\cal X}$, to denote sets. We write lowercase letters (e.g., $x$) to denote scalars. We use notation ${\bf{x}}$ to denote a vector and $x_i$ to denote the $i$-th element in ${\bf{x}}$. The bold letters, such as ${\bf{X}}$, denote matrices. Specially, the set of binary and real numbers are denoted by $\mathbb{B}$ and $\mathbb{R}$, respectively. In addition, we use the uppercase letter (e.g., $Y$) to denote random variable and the lowercase letter $y$ to represent a realization.

Throughout this paper, $\log \left(  \cdot  \right)$ denotes ``logarithm base 2'', and $\ln \left(  \cdot  \right)$ stands for the ``natural logarithm base $\rm e$'', where the constant ${\rm e} = 2.71828\dots$.

\subsection{Iterative Decoding Algorithms}

As described by in \cite{tanner}, \emph{Tanner graph} provides a complete representation of LDPC code and it aids in the description of decoding algorithms. A Tanner graph is a bipartite graph, that defines two sets of nodes: variable nodes (VNs) and check nodes (CNs). Each of the bits in the codeword corresponds to a VN, and each of the parity-check equations (rows of the parity-check matrix) corresponds to a CN. If a bit participates in a parity-check equation, there is an edge between the corresponding VN and CN.

Iterative decoding algorithms are operated on the Tanner graph \cite{Lin_shu_channel_coding}, which is a graphical representation of some parity check matrix that describes the code. The commonly used message in iterative decoding is the bit log-likelihood ratio (LLR). Given a noisy received signal vector $\mathbf y$ corresponding to a transmitted codeword $\mathbf x$, the LLR of the $v$-th bit in $\mathbf x$ is defined as
\begin{equation}\label{eq_LLR_def}
  {\ell_v} = \ln \frac{{\Pr \left( {{y_v}\left| {{x_v} = 0} \right.} \right)}}{{\Pr \left( {{y_v}\left| {{x_v} = 1} \right.} \right)}},
\end{equation}
where the $\Pr \left( {{y_v}\left| {{x_v}} \right.} \right)$ denotes the transition probability from $x_v$ to $y_v$.

The decoding of LDPC codes can be described as iterative message exchanges between VNs and CNs. During iteration $i$, the message passing from VN $v$ to CN $c$ is
\begin{equation}\label{eq_BP_v_c}
  \ell_{v \to c}^{\left( i \right)} = {\ell_v} + \sum\limits_{c^\prime \in {\mathcal N}\left( v \right)\backslash c} {\ell_{c^\prime \to v}^{\left( {i - 1} \right)}},
\end{equation}
where ${\mathcal N}\left( v \right)$ denotes the neighboring node set of $v$ consisting of CNs which are adjacent to the VN $v$, and ${\mathcal N}\left( v \right)\backslash c$ denotes the neighboring node set except the CN $c$.

The message passing from CN $c$ to VN $v$ is
\begin{equation}\label{eq_BP_c_v}
  \ell_{c \to v}^{\left( i \right)} = 2{\tanh ^{ - 1}}\left( {\prod\limits_{v^\prime \in {\mathcal N}\left( c \right)\backslash v} {\tanh \left( {\frac{{\ell_{v^\prime \to c}^{\left( i \right)}}}{2}} \right)} } \right),
\end{equation}
where ${\mathcal N}\left( c \right)$ denotes the neighboring node set of $c$ consisting of VNs which are adjacent to the CN $c$, and ${\mathcal N}\left( c \right)\backslash v$ denotes the neighboring node set except the VN $v$. At the first iteration, $\ell_{c \to v}^{\left( 0 \right)}$ is initialized to 0 in (\ref{eq_BP_v_c}).

After $i$ iterations, the soft estimation $s_v$ about the LLR for code bit $x_v$ is written as
\begin{equation}\label{eq_BP_final_LLR}
  {s_v} = {\ell_v} + \sum\limits_{c^\prime \in {\mathcal N}\left( v \right)} {\ell_{c^\prime \to v}^{\left( i \right)}},
\end{equation}
and a decision is made by
\begin{equation}\label{eq_BP_decision}
  {{\hat x}_v} = \frac{{1 - {\mathop{\rm sgn}} \left( {{s_v}} \right)}}{2},
\end{equation}
where the function ${\rm sgn}\left( s_v \right)$ denotes the sign of $s_v$.

The iterative decoding procedure described above is called the sum-product (SP) algorithm or the belief-propagation (BP) algorithm. To implement equation (\ref{eq_BP_c_v}), the SP decoder involves hyperbolic tangent functions and many multiplications. To reduce computational complexity, the min-sum (MS) algorithm \cite{minsum} is used to approximate equation (\ref{eq_BP_c_v}) as
\begin{equation}\label{eq_minsum_c_v}
  \ell_{c \to v}^{\left( i \right)}=\left( {\prod\limits_{v^\prime \in {\mathcal N}\left( c \right)\backslash v} {{\mathop{\rm sgn}} \left( {\ell_{v^\prime \to c}^{\left( i \right)}} \right)} } \right) \times \mathop {\min }\limits_{v^\prime \in {\mathcal N}\left( c \right)\backslash v} \left| {\ell_{v^\prime \to c}^{\left( i \right)}} \right|.
\end{equation}

Compared to the SP algorithm, the MS approximation suffers from non-negligible performance loss. Several enhanced MS algorithms have been proposed in \cite{minsum_variants}: the normalized min-sum (NMS) adds a normalization scaling factor $\alpha$ to (\ref{eq_minsum_c_v}) as
\begin{equation}\label{eq_normalized_minsum_c_v}
  \ell_{c \to v}^{\left( i \right)} = \alpha \times \left( {\prod\limits_{v^\prime \in {\mathcal N}\left( c \right)\backslash v} {{\mathop{\rm sgn}} \left( {\ell_{v^\prime \to c}^{\left( i \right)}} \right)} } \right) \times \mathop {\min }\limits_{v^\prime \in {\mathcal N}\left( c \right)\backslash v} \left| {\ell_{v^\prime \to c}^{\left( i \right)}} \right|.
\end{equation}
Another technique named the offset min-sum (OMS) adds an offset correction term to (\ref{eq_minsum_c_v}) as
\begin{equation}\label{eq_offset_minsum_c_v}
\begin{aligned}
\ell_{c \to v}^{\left( i \right)} = & \left( {\prod\limits_{v^\prime \in {\cal N}\left( c \right)\backslash v} {{\mathop{\rm sgn}} \left( {\ell_{v^\prime \to c}^{\left( i \right)}} \right)} } \right) \times \\
~ & \max \left\{ {\mathop {\min }\limits_{v^\prime \in {\cal N}\left( c \right)\backslash v} \left| {\ell_{v^\prime \to c}^{\left( i \right)}} \right| - \beta ,0} \right\}.
\end{aligned}
\end{equation}
The OMS method avoids multiplications, making it more suitable for practical iterative decoder implementation. For NMS and OMS algorithms, the critical $\alpha$ and $\beta$ should be optimized for every specific LDPC code by using the density evolution (DE) \cite{minsum_DE} and the recommended values are $\alpha = 0.8$, $\beta = 0.15$ in \cite{minsum_variants}\footnote{The normalizing factor in \cite{minsum_variants} is written as the division form, hence, the recommended value $1.25$ in \cite{minsum_variants} is converted to $\alpha = 1/1.25 = 0.8$ in this paper.} for some regular LDPC codes. They are viewed as one benchmark for comparison in many existing works \cite{minsum_DE,minsum_adaptive}. Thus, in this paper, we also adopt the NMS decoding with $\alpha = 0.8$ and the OMS decoding with $\beta = 0.15$ as two baseline algorithms. Later, although some adaptive methods \cite{minsum_adaptive} were proposed to recognize the signal amplitude and noise variance, they also increase the computational complexity. Hence, there are still no analytical results for the choice of $\alpha$ and $\beta$, and the current optimization methods can only deal with single parameter cases.

\subsection{Protograph LDPC Codes}

\begin{figure}[t]
\setlength{\abovecaptionskip}{0.cm}
\setlength{\belowcaptionskip}{-0.cm}
  \centering{\includegraphics[scale=0.8]{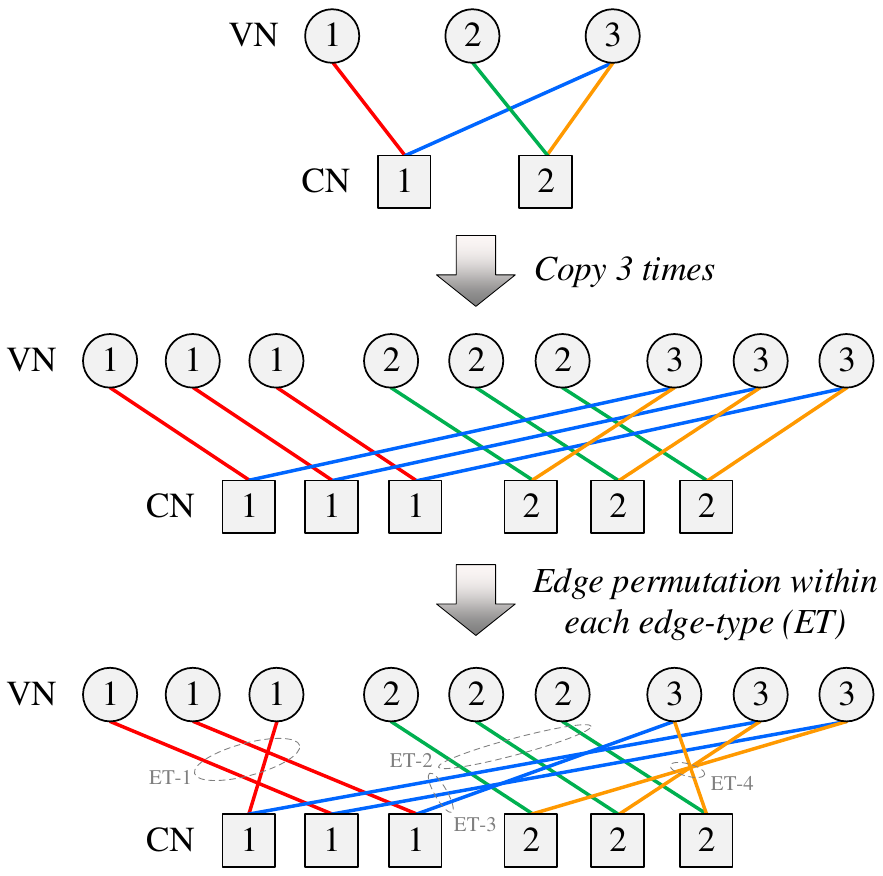}}
  \caption{A graphical demonstration of protograph LDPC codes.}\label{fig_protograph_LDPC}
\end{figure}

Multi-edge type LDPC (MET-LDPC) codes \cite{muti_edge_LDPC} define a unified framework of structured LDPC codes. Unlike the Tanner graph of conventional LDPC codes with only a single edge-type, MET-LDPC codes provide multiple edge-types and allow exploring better codes that are optimized under specific constraints. These constraints are usually designed to improve the efficiency of encoding and decoding.

Protograph LDPC codes \cite{protograph_LDPC} are a subclass of MET-LDPC codes, which is defined using a small base code. The LDPC codes adopted in 5G NR are the protograph codes \cite{5G_NR_std}, and Fig. \ref{fig_protograph_LDPC} shows a toy example of protograph LDPC codes. A set of edge-types are defined in an $\left( N_b,K_b\right)$ base code, where $N_b$ and $N_b - K_b$ equal the number of VNs and CNs in the Tanner graph corresponding to this base code, i.e., the base graph. According to the connections to these edges, variable and check node-types are determined, respectively. An $\left( N,K\right)$ code is derived by taking $Z = N/{N_b} = K/{K_b}$ replicas of the base graph and then permuting the edges within the same edge-type to integrate the separated $Z$ base graph replicas into a larger one. This operation is called \emph{lifting}, and the derived code is also named the ``$Z$-lifted'' LDPC code. Since the edge permutation is performed only within a bundle of edges of each edge-type, the structural properties of edges and nodes (e.g., degree) in the lifted codes remain the same with that in the base code. Hence, the macroscopic structure of a protograph LDPC code can be captured by its small base graph \cite{5G_NR_std_magazine}. This encapsulation property of protograph LDPC codes allows our proposed neural MS decoder to apply the same correction term to all edges of the same edge-type. The details will be presented in Section \ref{section_neural_min_sum}.

\section{The Proposed Neural Min-Sum Decoder}\label{section_neural_min_sum}

In this section, we present details of the optimized MS decoding by using customized sparse neural networks.

\subsection{Neural MS Decoding for Protograph LDPC Codes}

The above normalizing and offset factors can be viewed as the assigned weights and biases to the edges in the Tanner graph. Thus, to find the optimal parameters, a straightforward method is to construct an MS decoding neural network to automatically learn these parameters. To this end, according to the Tanner graph, we can construct a sparse neural decoder with a not-fully connected structure that is a trellis representation of iterative decoding. The neural network input layer is a vector of size $N$, which is the code block length (i.e., the number of variable nodes in the Tanner graph). All the subsequent layers in the trellis, except for the last one (i.e., all the hidden layers), are of size $E$, where $E$ denotes the number of edges in the Tanner graph. For ease of exposition, we mark the hidden layer index $i$ corresponding to the $i$-th iteration in the MS decoding process. Each hidden layer includes two sublayers $i_v$ and $i_c$ corresponding to VN update and CN update, respectively. The output layer contains $N$ neurons.

For hidden layer $i$, each \emph{processing element (PE)} in sublayer $i_v$ outputs the message along the edge sent from the associated VN to CN, and each \emph{neuron} in sublayer $i_c$ outputs the message along the edge sent from the associated CN to VN. The PE in the first hidden layer (sublayer $1_v$) corresponding to the edge $e = \left(v,c\right)$ is connected to the $v$-th element in the input layer. The PEs in hidden layer $i$ ($i>1$) (sublayer $i_v$) corresponding to the edge $e = \left(v,c\right)$ is connected to the neurons in hidden layer $\left(i-1\right)$ (sublayer $\left(i-1\right)_c$) associated with the edges $e' = \left(v,c'\right)$ for $c' \neq c$. The neurons in hidden layer $i$ ($i\ge1$) (sublayer $i_c$) corresponding to the edge $e = \left(v,c\right)$ is connected to the PEs in sublayer $i_v$ associated with the edges $e' = \left(v',c\right)$ for $v' \neq v$.

\begin{figure}[t]
\setlength{\abovecaptionskip}{0.cm}
\setlength{\belowcaptionskip}{-0.cm}
  \centering{\includegraphics[scale=0.64]{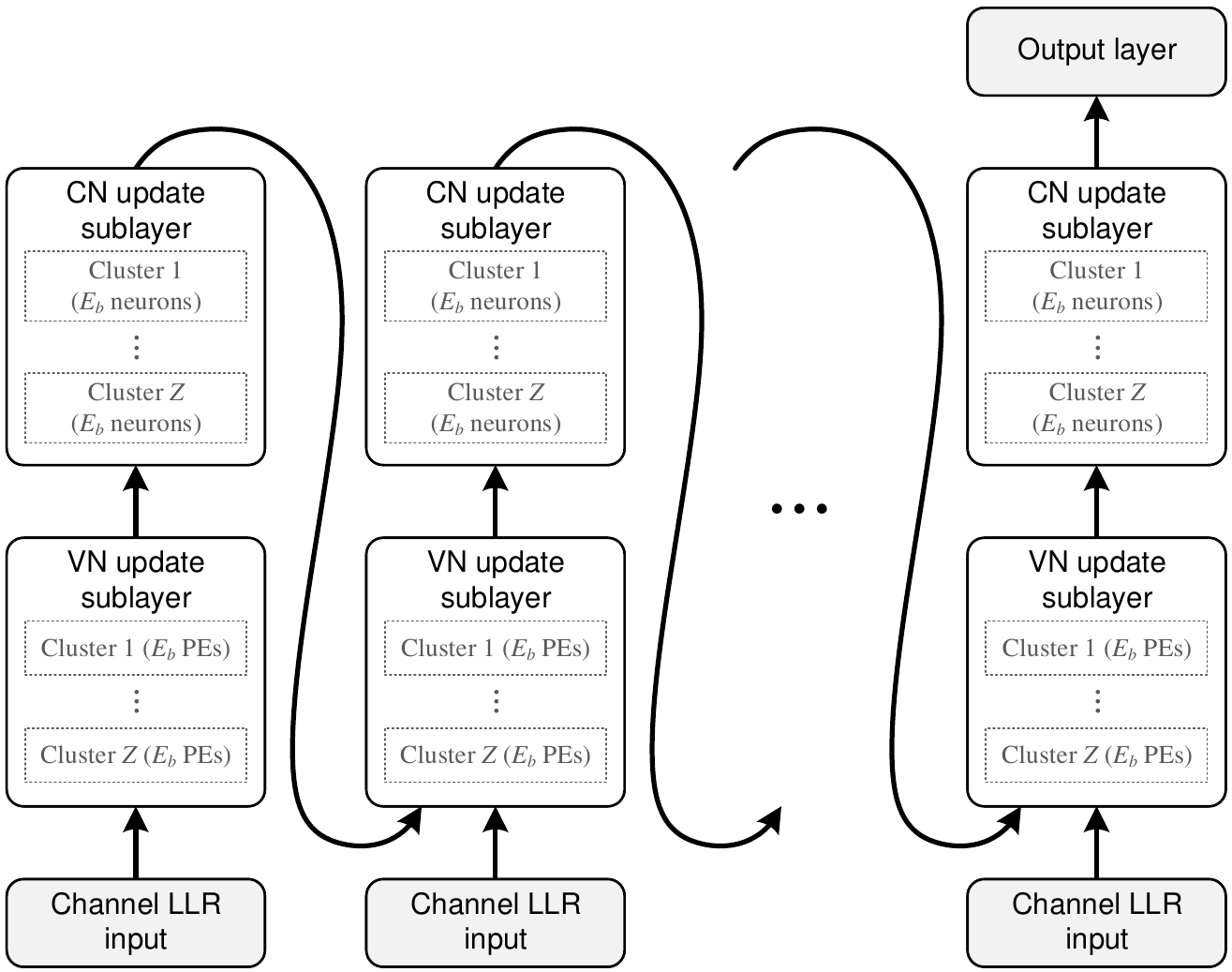}}
  \caption{Network structure of the proposed neural MS decoder.}\label{fig_NN}
\end{figure}

Regarding the protograph LDPC codes, all the $E$ PEs or neurons in each hidden layer can be divided to $Z$ clusters, each cluster contains $E_b$ elements corresponding to the number of edges on the base graph, thus we have $E = ZE_b$. Given an $\left( N_b,K_b\right)$ base code $C_b$, for the $Z$-lifted code, we define the set ${\mathcal E}_{e_b}$ denoting the \emph{$Z$-bundle of edges} derived from $e_b = \left(v_b,c_b\right)$ as
\begin{equation}\label{eq_ET_set}
\begin{aligned}
  & {\mathcal E}_{e_b = \left(v_b,c_b\right)} \triangleq \\
  & \left\{ {e = \left( {v,c} \right)\left| \begin{aligned}
& v = {v_b} + {\lambda _v}{N_b},{\lambda _v} \in \left[\kern-0.15em\left[ {N - 1}
 \right]\kern-0.15em\right],\\
& c = {c_b} + {\lambda _c}\left( {{N_b} - {K_b}} \right),{\lambda _c} \in \left[\kern-0.15em\left[ {N - 1}
 \right]\kern-0.15em\right]
\end{aligned} \right.} \right\},
\end{aligned}
\end{equation}
where the set $\left[\kern-0.15em\left[ {N - 1} \right]\kern-0.15em\right] \triangleq \left\{0,1,\cdots,N-1\right\}$. Thus, the PEs or neurons in each cluster belong to a different set ${\mathcal E}_{e_b = \left(v_b,c_b\right)}$. A demonstration of the neural MS decoder is given in Fig. \ref{fig_NN}.

\begin{table*}[t]
\renewcommand{\arraystretch}{1.6}
  \centering
  \small
  \caption{Four types of neural MS decoder.}\label{table_neural_MS}
  \begin{tabular}{!{\vrule width1pt}m{1.3cm}|m{4cm}|m{9.5cm}!{\vrule width1pt}}
    \Xhline{1pt}
    \centering Type & \centering Description & \centering Definition ($i'$ and $i''$ denote two different iterations) \tabularnewline
    \Xhline{1pt}
    \centering Type-I & \centering neural NOMS & \centering ${\bm \alpha}^{(i')} \neq {\bm \alpha}^{(i'')}$, ${\bm \beta}^{(i')} \neq {\bm \beta}^{(i'')}$ with $i' \neq i''$ \tabularnewline
    \hline
     \centering Type-II & \centering simplified neural NOMS & \centering ${\bm \alpha}^{(i)} = {\alpha}^{(i)}$, ${\bm \beta}^{(i)} = { \beta}^{(i)}$, ${ \alpha}^{(i')} \neq { \alpha}^{(i'')}$, ${ \beta}^{(i')} \neq { \beta}^{(i'')}$ with $i' \neq i''$ \tabularnewline
     \hline
     \centering Type-III & \centering simplified neural NMS & \centering ${\bm \alpha}^{(i)} = {\alpha}^{(i)}$, ${\bm \beta}^{(i)} = 0$, ${ \alpha}^{(i')} \neq { \alpha}^{(i'')}$ with $i' \neq i''$ \tabularnewline
     \hline
     \centering Type-IV & \centering simplified neural OMS & \centering ${\bm \alpha}^{(i)} = 1$, ${\bm \beta}^{(i)} = { \beta}^{(i)}$, ${ \beta}^{(i')} \neq { \beta}^{(i'')}$ with $i' \neq i''$ \tabularnewline
     \Xhline{1pt}
  \end{tabular}
\end{table*}

The messages transmitted over the neural MS decoder are as following. Given the number of iterations $I$, consider the $i$-th hidden layer, $i=1,2,\cdots,I$, and the $e = \left(v,c\right)$ is the index of some PE in sublayer $i_v$. The output message of this PE is written as
\begin{equation}\label{eq_neural_VN}
  \ell_{e = \left(v,c\right)}^{\left( i_v \right)} = {\ell_v} + \sum\limits_{e^\prime = \left(v,c^\prime\right),c^\prime \neq c} {\ell_{e^\prime}^{\left({\left( {i - 1} \right)_c}\right)}},
\end{equation}
where ${\ell_{e^\prime}^{\left( {0} \right)}} = 0$ for all $e^\prime$ at the initialization step. Regarding the protograph structure, for any $e \in {\mathcal E}_{e_b = \left(v_b,c_b\right)}$, we can derive that $e^\prime \in {\mathcal E}_{e_b^\prime = \left(v_b,c_b^\prime\right)}$ with $c_b^\prime \neq c_b$ in (\ref{eq_neural_VN}). The output message of neuron $e = \left(v,c\right)$ in the $i_c$-th sublayer is written as
\begin{equation}\label{eq_neural_CN}
\begin{aligned}
  \ell_{e = \left(v,c\right)}^{\left( i_c \right)} = & \left( {{\prod _{e^\prime = \left(v^\prime,c\right),v^\prime \neq v}}{\rm{sgn}}\left( {\ell_{e^\prime}^{\left( i_v \right)}} \right)} \right) \times \\
  ~ & {\text{ReLU}}\left( {\alpha_{e}^{\left( i \right)} \times \mathop {\min }\limits_{e^\prime = \left(v^\prime,c\right),v^\prime \neq v} \left| {\ell_{e^\prime}^{\left( i_v \right)}} \right| - \beta_{e}^{\left( i \right)}} \right),
  \end{aligned}
\end{equation}
where ${\text{ReLU}}\left( \cdot \right)$ is one of the most commonly used activation functions in deep learning studies,
\begin{equation}
  {\text{ReLU}}\left( x \right) = \max\left( x,0\right).
\end{equation}
Clearly, different from previous works \cite{deep_learning_decoding_Allerton,deep_learning_decoding_ISIT,deep_learning_decoding_JSTSP}, where \cite{deep_learning_decoding_Allerton} trained a weighted SP decoder and \cite{deep_learning_decoding_ISIT} trained an OMS decoder with learnable offsets, we add both normalizing factors and offset factors to (\ref{eq_minsum_c_v}) of the MS algorithm, i.e., the equation (\ref{eq_neural_CN}). This design is indeed consistent with classical neural networks \cite{deep_learning_chenmingzhe1} equipped with both weights and biases so that the degrees of freedom for optimization are expanded. Compared to traditional neural decoders with only weights or biases, the proposed neural MS decoding can better compensate for the min-sum loss, and partially mitigate the message correlation due to short cycles in the Tanner graph as a bonus. The output neuron gives the information
\begin{equation}\label{eq_neural_output}
  {o_v} = \sigma \left( {{\ell _v} + \sum\limits_{e^\prime = \left( {v,c^\prime} \right)} {\ell _{e^\prime}^{\left( {{I_c}} \right)}} } \right),
\end{equation}
where $\sigma \left( x \right) = {\left( {1 + \exp \left( { - x} \right)} \right)^{ - 1}}$ is the sigmoid function, it ensures the final network output is in the range $\left[0,1\right]$ denoting the probability of transmitted bit $x_v = 0$. Also, note that by setting all the weights ${\alpha}_{e}^{\left(i\right)}$ to 1 and all the biases ${\beta}_{e}^{\left(i\right)}$ to 0, the proposed neural MS decoder regenerates the standard MS decoder as (\ref{eq_BP_v_c}) and (\ref{eq_minsum_c_v}).

The weights $\left\{{\alpha}_{e=\left(v,c\right)}^{\left(i\right)}\right\}$ and biases $\left\{{\beta}_{e=\left(v,c\right)}^{\left(i\right)}\right\}$ within neurons vary for different
iteration $i$ for $i=1,2,\cdots,I$. The message updating (\ref{eq_BP_v_c}) from VNs to CNs remains unchanged which is completed within the PEs as (\ref{eq_neural_VN}). The proposed neural MS decoding method can be easily extended to the neural SP decoding method by changing the message update rule in \eqref{eq_neural_CN} as follows:
\begin{equation}\label{eq_neural_SP_CN}
\begin{aligned}
  & \ell _{e = \left( {v,c} \right)}^{\left( {{i_c}} \right)} = \\
  & \alpha _e^{\left( i \right)} \times 2{\tanh ^{ - 1}}\left( {\prod\limits_{e' = \left( {v',c} \right),v' \ne v} {\tanh \left( {\frac{{\ell _{e'}^{\left( {{i_v}} \right)}}}{2}} \right)} } \right) + \beta _e^{\left( i \right)}.
\end{aligned}
\end{equation}
Hereinafter, we focus on the neural MS decoding method and the neural SP decoding method is set as a comparison in Section \ref{section_performance_evaluation}.

\begin{proposition}[parameter-sharing mechanism]\label{prop_training_constrain}
\emph{
  The macroscopic structure of a protograph LDPC code can be captured by its small base graph \cite{5G_NR_std_magazine}. Taking advantage of this encapsulation property of protograph LDPC codes, an additional restriction is applied to the weights and biases: for any edge pair $e_1$ and $e_2$ belonging to the same edge-type, i.e., ${e_1} \in {\mathcal E}_b$ and ${e_2} \in {\mathcal E}_b$, the same values are applied to the neurons corresponding to these two edges, i.e.,
  \begin{subequations}
    \begin{equation}
      {\alpha}_{e_1}^{\left( i\right)} = {\alpha}_{e_2}^{\left( i\right)},
    \end{equation}
    \begin{equation}
      {\beta}_{e_1}^{\left( i\right)} = {\beta}_{e_2}^{\left( i\right)}.
    \end{equation}
  \end{subequations}
  Moreover, one parameter array may be applied to multiple lifted codes derived from the same base code.
}
\end{proposition}

With the protograph-based parameter-sharing mechanism in \emph{Proposition \ref{prop_training_constrain}}, to decode a set of protograph LDPC codes derived from the same base code, only a small weight/bias array must be adapted. In particular, the size of the parameter array is proportional to the number of edge-types, or equals the number of edges in the base graph\footnote{If the base graph has parallel edges as in MET-LDPC codes \cite{muti_edge_LDPC}, the conclusion is still valid since these parallel edges deliver the same LLR message so that they can share the same parameters.}. Hence, compared to applying independent correction terms to every edge in the Tanner graph of every specific code \cite{deep_learning_decoding_Allerton,deep_learning_decoding_ISIT,deep_learning_decoding_JSTSP}, the proposed protograph-based parameter-sharing mechanism can efficiently reduce training complexity and memory cost. Furthermore, it breaks the limitation of codelength, thus the proposed neural MS decoder can be applied to moderate and long codelengths, which is desired for practical use.

In practical implementation, the proposed neural MS decoder includes four basic types summarized in Table \ref{table_neural_MS}. For iteration $i$, the parameter array ${\bm \alpha}^{\left(i\right)}$ includes $E_b$ elements ${\alpha}_{e}^{\left(i\right)}$ and each one represents a different edge-type. Also, the parameter array ${\bm \beta}^{\left(i\right)}$ includes $E_b$ elements ${\beta}_{e}^{\left(i\right)}$. In fact, neural MS decoding is an integrated version of NMS and OMS algorithms with fine-tuned factors so that it is named ``neural normalized\&offset MS (neural NOMS)''. To reduce complexity, we tie the tuned factors and introduce three simplified versions of the neural NOMS. As shown in Table \ref{table_neural_MS}, the Type-II decoder sets the normalizing and offset factors within one iteration as two paramters and they vary as the number of iterations increases. Type-III and Type-IV are the simplified neural NMS and OMS decoding methods, respectively, and have been studied in previous work \cite{deep_learning_decoding_ISIT_simplfied}.

\subsection{Training the Neural MS Decoder}

In this paper, we use the expected cross-entropy between the transmitted codeword $\mathbf x$ and the neural MS decoder output $\mathbf o$ for the loss function, which has been proven to be effective in previous works \cite{deep_learning_decoding_Allerton,deep_learning_decoding_ISIT,deep_learning_decoding_JSTSP}, defined as
\begin{equation}\label{eq_neural_loss_func}
  L\left( {{\mathbf o},{\mathbf x}} \right) =  - \frac{1}{N}\sum\limits_{v = 1}^N {{x_v}\log \left( {{o_v}} \right) + } \left( {1 - {x_v}} \right)\log \left( {1 - {o_v}} \right),
\end{equation}
where $o_v$ and $x_v$ denote the neural network output and the $v$-th element of the transmitted codeword.

Given a particular normalizing and offset vector $\bm \alpha$ and $\bm \beta$, the loss function (\ref{eq_neural_loss_func}) can be estimated. By computing the gradient ${\nabla _{{\bm \alpha},{\bm \beta}}}L$, the normalizing/offset factors can be tuned by gradient-based optimization methods which are widely used in deep learning
studies, e.g., ADAM \cite{adam}.

To find the optimal normalizing/offset factors, a straightforward training method is to construct an MS decoding neural network according to (\ref{eq_neural_VN}), (\ref{eq_neural_CN}) and (\ref{eq_neural_output}), and tuning the weights and biases for all the $I$ iterations, i.e., $\left\{{\bm \alpha}^{\left(i\right)}\right\}$ and  $\left\{{\bm \beta}^{\left(i\right)}\right\}$ with $i=1,2,\cdots,I$, to minimize
the loss function in (\ref{eq_neural_loss_func}). The gradients are propagated from
the network layer of the $I$-th iteration to that of the first iteration. With a practical iteration number $I$, e.g., 25 or 50, the network structure is quite deep and suffers from the \emph{gradient vanishing problem} when training. In addition, the generalization ability of one trained neural MS decoder to various codelengths and code rates is a key for practical use. To address these two problems, we propose two training methods as follows:

\subsubsection{Codelength/Rate Compatible Training}

We do not restrict the training samples selected from some one particular code, even though it can achieve the best performance for this code. According to \emph{Proposition \ref{prop_training_constrain}}, the same parameter array can be applied to multiple codes derived from the same base graph. Therefore, to enhance the generalization ability of one trained neural MS decoder to multiple codelengths and code rates, we can train the neural network by randomly selecting samples from a set of codes ${\mathcal C} = \left\{C_1,C_2,\cdots\right\}$ with different lengths or rates, and these codes are derived from one identical base graph $C_b$. In this way, one trained neural MS decoder can match multiple codes.

\subsubsection{Iteration-by-Iteration Greedy Training}

To address the vanishing gradient problem in a deep neural decoder, \cite{deep_learning_decoding_Allerton,deep_learning_decoding_ISIT,deep_learning_decoding_JSTSP} employ a multi-loss function to inject the gradients directly to every layer. In this paper, we propose a different way: \emph{the network is built and trained iteration-by-iteration greedily.} Once a layer is trained, the corresponding weights and biases are fixed in the training for later iterations. The neural decoding network is growing from a one-layer network for only one iteration to a multi-layer network for $I$ iterations; each time, only the weights/biases in the last layer are learnable.

Compared to the multi-loss training method in \cite{deep_learning_decoding_JSTSP}, one would expect a performance loss under this greedy training method. However, in practice, the number of iterations used for decoding cannot be very large, i.e., $I \le 50$, the potential performance loss is negligible. Although the value of $I$ is moderate for the greedy strategy, the corresponding neural decoding network that consists of $I$, e.g., 25 or 50, layers has already been quite deep, which may lead to the vanishing gradient problem when directly training all $I$ layers.

\begin{figure}[t]
\setlength{\abovecaptionskip}{0.cm}
\setlength{\belowcaptionskip}{-0.cm}
  \centering{\includegraphics[scale=0.9]{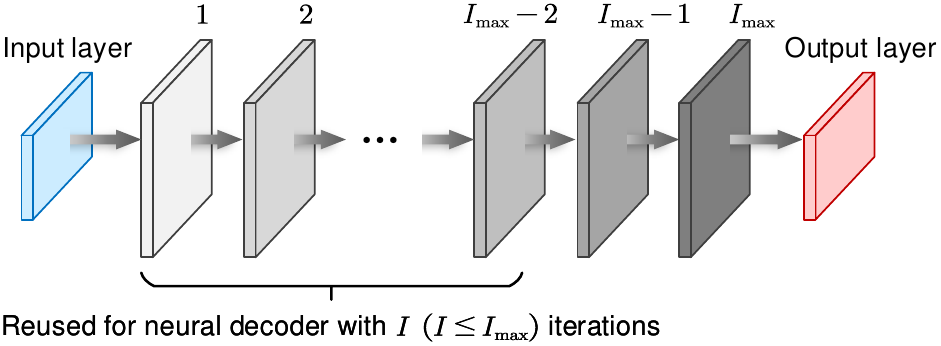}}
  \caption{A sketch of the reusability of hidden layers trained with the proposed iteration-by-iteration manner.}\label{fig_NN_reuse}
\end{figure}

In summary, the proposed greedy training method has the following advantages:
\begin{itemize}
  \item The actual network for training always has a shallow structure so as to avoid the vanishing gradient problem and reduce the training complexity.

  \item Per-edge-type parameters and the greedy training method enable the neural MS decoder to mitigate the message correlation due to short cycles in the Tanner graph and tend to optimize the performance of neural decoders as early as possible, i.e., faster convergence, especially for some short codelength cases. This is consistent with the early termination mechanism widely-used in LDPC iterative decoding.

  \item The one-iteration training granularity enables the neural MS decoder to flexibly set up its decoding configurations, i.e., the number of iterations. In this way, many parameters in the optimized MS decoder can be reused rather than retrained for every different configuration. As shown in Fig. \ref{fig_NN_reuse}, the neural network is trained for $I_{\max}$ iterations, then for any neural decoder with $I$ ($I\le I_{\max}$) iterations, it can be directly built by recycling the first $I$ hidden layers in the trained neural decoder.
\end{itemize}

\begin{algorithm}[htbp]
\setlength{\abovecaptionskip}{0cm}
\setlength{\belowcaptionskip}{-0cm}
\caption{Training the neural MS decoder}\label{alg_traning}
\KwIn {The base code $C_b$, the maximum iteration numbers $I_{\max}$, a set of lifted codes used for training ${\mathcal C} = \left\{C_1,C_2,\cdots\right\}$ derived from $C_b$, and the SNR table where ${\mathsf{SNR}}\left(i,j\right)$ denotes the training SNR for the $i$-th iteration of code $C_j$\;}

\For{$k = 1,2,\cdots,I_{\max}$}
{
    \Comment{~iteration-by-iteration training}
    Initialize a set of neural MS decoders with shared weights and biases $\left\{{\bm \alpha}^{\left(i\right)}\right\}$ and $\left\{{\bm \beta}^{\left(i\right)}\right\}$ with iteration $i=1,2,\cdots,k$ which correspond to all $C_j \in {\mathcal C}$, respectively\;

    Weight/bias vectors for iteration $i=1,2\cdots,k-1$ are initialized to previously learned values\;

    Initialize ${\bm \alpha}^{\left(k\right)}$ and ${\bm \beta}^{\left(k\right)}$ corresponding to the last iteration with random values\;

    \Repeat{converge {\bf or} reach a maximum step number}
    {
        Randomly select a code $C_j \in {\mathcal C}$\;
        Randomly generate a codeword $\mathbf x$ of $C_j$\;
        Generate the received signal $\mathbf y$ by sending $\mathbf x$ through a BPSK modulated AWGN channel with ${\mathsf{SNR}}\left(k,j\right)$\;
        Feed $\mathbf y$ into the neural MS decoder corresponding to $C_j$ and obtain output $\mathbf s$\;
        Compute loss function according to (\ref{eq_neural_loss_func})\;
        Update ${\bm \alpha}^{\left(k\right)}$ and ${\bm \beta}^{\left(k\right)}$ using gradient descent algorithm\;
    }

}
\Return $\left\{{\bm \alpha}^{\left(i\right)}\right\}$ and $\left\{{\bm \beta}^{\left(i\right)}\right\}$ with $i=1,2,\cdots,I_{\max}$.\\
\end{algorithm}

Since the bit error rate (BER) could vary a lot with different iteration numbers, the SNR should be assigned to different levels during the training process. In addition, note that the weights/biases are shared among all the possible lifted codes from the same base code, to prevent overfitting to a specific code, the training set should include samples from a set of codes with multiple lifting factors $Z$. Therefore, a table of SNRs for multiple codes under different iteration numbers is required for training. These SNR values should be selected to have the same BER performance under a reference decoding algorithm, e.g., the SP algorithm, to make all the codes used for training are (approximately) fairly handled.

The proposed training methods for the neural MS decoder are summarized in Algorithm \ref{alg_traning}, which can be easily extended to a batch training variant.

\section{Learning to Damp}\label{section_damping}

\begin{table*}[t]
\renewcommand{\arraystretch}{1.6}
  \centering
  \small
  \caption{Two types of neural MS decoder with damping factors.}\label{table_damp_neural_MS}
  \begin{tabular}{!{\vrule width1pt}m{1.2cm}|m{5.5cm}|m{9.5cm}!{\vrule width1pt}}
    \Xhline{1pt}
    \centering Type & \centering Description & \centering Definition ($i'$ and $i''$ denote two different iterations) \tabularnewline
    \Xhline{1pt}
    \centering Type-V & \centering neural NOMS with damping & \centering ${\bm \alpha}^{(i')} \neq {\bm \alpha}^{(i'')}$, ${\bm \beta}^{(i')} \neq {\bm \beta}^{(i'')}$, ${\bm \gamma}^{(i')} \neq {\bm \gamma}^{(i'')}$ with $i' \neq i''$\tabularnewline
    \hline
     \centering Type-VI & \centering simplified neural NOMS with damping & \centering ${\bm \alpha}^{(i')} \neq {\bm \alpha}^{(i'')}$, ${\bm \beta}^{(i')} \neq {\bm \beta}^{(i'')}$, ${\bm \gamma}^{(i)} = {\gamma}^{(i)}$, ${ \gamma}^{(i')} \neq { \gamma}^{(i'')}$ with $i' \neq i''$ \tabularnewline
     \Xhline{1pt}
  \end{tabular}
\end{table*}

Apart from optimizing the output LLR messages from CN update, one additional way is to relax \cite{deep_learning_decoding_JSTSP} or damp \cite{damp} the output LLR messages from VN update. By introducing the damping factor, the convergence rate of a neural MS decoder can be further improved \cite{damp_origin}. In addition, the message correlation due to short cycles in the Tanner graph can be further mitigated, and thus the performance of the neural MS decoder for short codelengths is improved.

At iteration $i$, the message is damped by obtaining a convex combination of the message computed at iteration $(i-1)$ and the message at iteration $i$, with damping factors. Hence, in the neural MS decoder, the output message of the PE $e = \left(v,c\right)$ in sublayer $i_v$ is written as
\begin{equation}\label{eq_damp_VN}
\begin{aligned}
  \ell_{e = \left(v,c\right)}^{\left( i_v \right)} =  \gamma_{e}^{\left( i \right)}{\ell}_{e}^{\left( {\left( {i - 1} \right)_v} \right)} + \left(1 - \gamma_{e}^{\left( i \right)}\right){\tilde \ell}_{e}^{\left( i_v \right)},
\end{aligned}
\end{equation}
where
\begin{equation}
  {\tilde \ell}_{e}^{\left( i_v \right)} = {\ell_v} + \sum\limits_{e^\prime = \left(v,c^\prime\right),c^\prime \neq c} {\ell_{e^\prime}^{\left({\left( {i - 1} \right)_c}\right)}},
\end{equation}
and the damping factor $0 \le \gamma_{e}^{\left( i \right)} < 1$. Clearly, as $\gamma_{e}^{\left( i \right)} \to 0$, the decoder becomes less damped, and as $\gamma_{e}^{\left( i \right)} \to 1$, the decoder becomes more damped. When $\gamma_{e}^{\left( i \right)} = 0$, the decoder reverts to being a normal decoder.

\begin{figure}[t]
\setlength{\abovecaptionskip}{0.cm}
\setlength{\belowcaptionskip}{-0.cm}
  \centering{\includegraphics[scale=0.58]{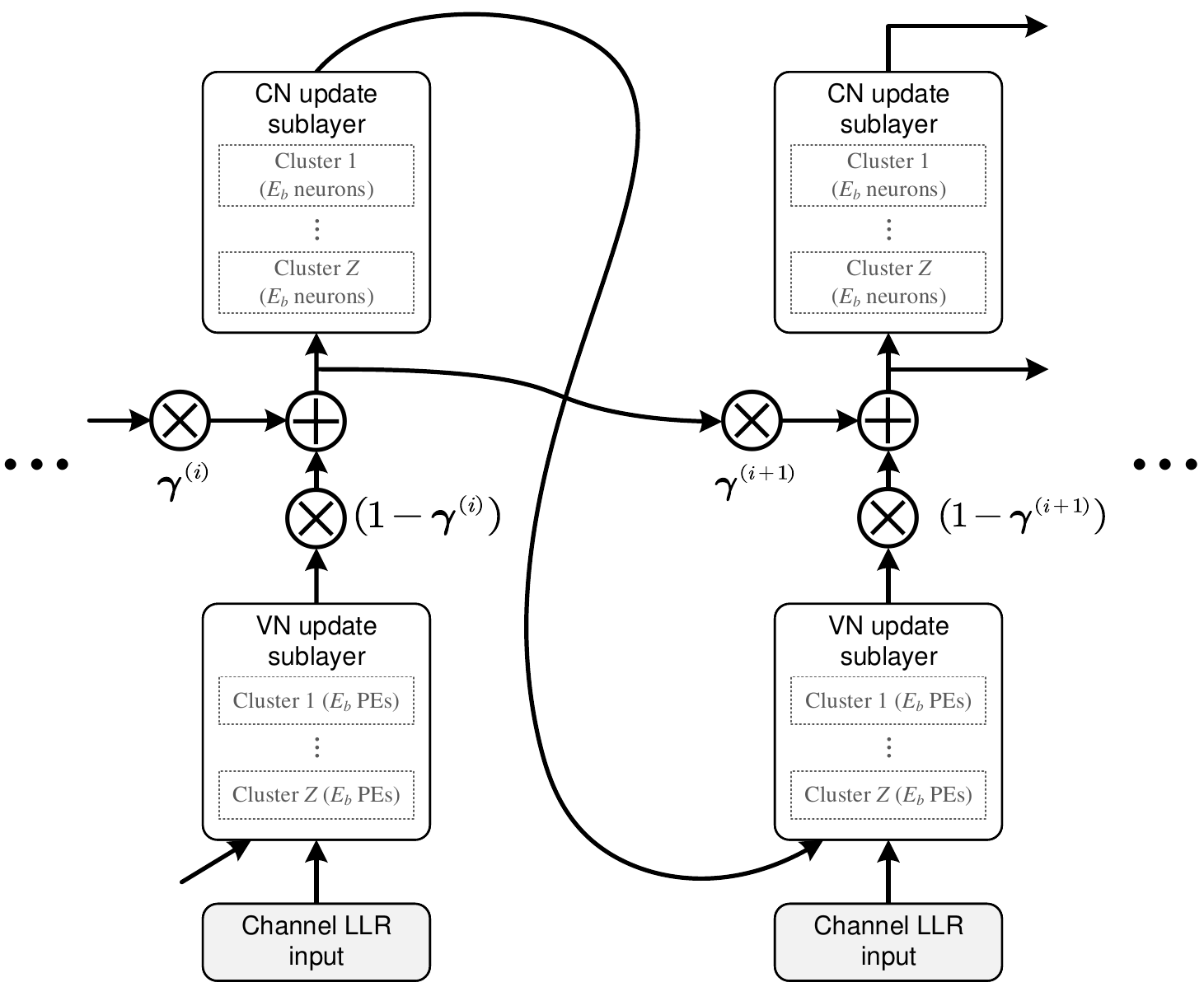}}
  \caption{Network structure of the neural MS decoder with damping factors.}\label{fig_NN_damp}
\end{figure}

Combining the damping factors, we further derive two types of neural MS decoders as Table \ref{table_damp_neural_MS} that follow the four basic types in Table \ref{table_neural_MS}. For the Type-V neural MS decoder, the per-edge damping factor array of each iteration ${\bm \gamma}^{\left(i\right)}$ also obeys the parameter-sharing mechanism in \emph{Proposition \ref{prop_training_constrain}}, i.e., the size of ${\bm \gamma}^{\left(i\right)}$ only equals the number of edge-types. During the training phase, we also adopt the iteration-by-iteration method to avoid the vanishing gradient problem. Also, under different numbers of iterations, a number of parameters in the optimized MS decoder could be reused rather than retrained for every different configuration. The structure of a damped neural MS decoder is figuratively given in Fig. \ref{fig_NN_damp}.

\section{T-EXIT Convergence Analysis}\label{section_T_EXIT}

In this section, we perform a trajectory-based EXIT (T-EXIT) convergence analysis for various decoders. EXIT chart \cite{brink2001,brink2004} is a powerful tool to analyze the convergence of iterative decoders. It tracks the mutual information (MI) transfer process during iterations. However, the conventional EXIT analysis cannot be roughly applied to protograph LDPC codes. Considering the different edge-connection properties, Liva \emph{et al}. proposed the protograph EXIT (PEXIT) algorithm \cite{liva2007}, where the MI is calculated for each VN and CN rather than degree-distribution pair. The PEXIT analysis provides a remarkably accurate and simple prediction of the decoding threshold for many code classes. Nevertheless, note that:
\begin{itemize}

  \item PEXIT is an asymptotic method used to determine the iterative decoding threshold for an LDPC decoding process assuming the number of decoding iterations is unlimited and the codelength is infinite. Since the number of iterations needed by the proposed neural MS decoder is limited and the codelength is finite, we cannot apply the PEXIT method to determine the decoding threshold and predict the asymptotic performance of the decoding process of the proposed decoder.

  \item PEXIT can only be appropriately applied to SP decoding under the AWGN channel \cite{liva2007}. For other algorithms, e.g., MS and its variants, the critical duality relationship \cite{Lin_shu_channel_coding} between VN and CN cannot be accurately satisfied. Consequently, the MI update at CNs cannot be analytically calculated. Meanwhile, the updated LLRs at CNs are not Gaussian so that the MI is hard to track.

\end{itemize}

Recently, a scattered EXIT (S-EXIT) chart method \cite{brink2018} was developed to optimize the degree profiles of short LDPC codes. By simulating several individual instances of a short code, one can obtain corresponding EXIT trajectories of an actual iterative decoder, and the vertices of these trajectories form a clutter. This S-EXIT method utilizes the statistics of numerous EXIT trajectories, and tracks their frequency of occurrence over the EXIT MI plane, thus enabling us to gain insight into the iterative decoding behavior. For the proposed decoder, there is no need for us to ``see through the clutter'' empirically as in S-EXIT since we only care about the average reliability of propagated messages.

To evaluate the convergence performance of an iterative decoder, we track the average MI (AMI) transfer process between VNs and CNs such that we can provide a virtual representation of the iterative decoding process. This trajectory-based EXIT method is named T-EXIT.

For iteration $i$, we denote $I_{A,\rm{VN}}^{\left(i\right)}$ ($I_{A,\rm{CN}}^{\left(i\right)}$) as the \emph{a priori} AMI between input LLRs $\ell_{c \to v}^{\left(i-1\right)}$ ($\ell_{v \to c}^{\left(i\right)}$) and the corresponding code bits. Similarly, we denote $I_{E,\rm{VN}}^{\left(i\right)}$ ($I_{E,\rm{CN}}^{\left(i\right)}$) as the extrinsic AMI between output LLRs $\ell_{v \to c}^{\left(i\right)}$ ($\ell_{c \to v}^{\left(i\right)}$) and the code bits. The target of T-EXIT analysis is to obtain two transfer functions
\begin{subequations}\label{eq_EXIT_functions}
\begin{equation}
  I_{E,\rm{VN}}^{\left(i\right)} = T_v\left(I_{A,\rm{VN}}^{\left(i\right)}\right),
\end{equation}
\begin{equation}
  I_{E,\rm{CN}}^{\left(i\right)} = T_c\left(I_{A,\rm{CN}}^{\left(i\right)}\right).
\end{equation}
\end{subequations}
To this end, we first collect all the output LLRs of VNs and CNs at each iteration by using a number of simulated code blocks. Then, we count the probability density functions (PDFs) of these LLRs with the histogram method, and calculate the AMI as
\begin{equation}\label{AMI}
I_Y = I\left(X;Y\right) = \sum\limits_{x\in{\mathbb{B}}}\int_{\mathbb{R}}p_Y(y\vert x)p_X\left(x\right)\log\frac{p_Y\left(y\vert x\right)}{p_Y\left(y\right)}{\rm d}y.
\end{equation}
In this way, we can get a zigzag path that reflects the decoding trajectory. Then, by connecting the vertices at both ends of the zigzag path separately, we can get two curves corresponding to $T_v\left(\cdot\right)$ and $T_c^{-1}\left(\cdot\right)$, respectively. The specific steps of T-EXIT are summarized as following.

\begin{itemize}
  \item \textbf{Initialization}

  Initialize the LDPC code $C$, the number of iterations $I$, the SNR, and the number of simulated blocks $K$ ($K$ is sufficiently large);

  Initialize $I_{A,\rm{VN}}^{\left(1\right)}=I_{E,\rm{CN}}^{\left(0\right)}=0$;

  Initialize vectors ${\bm \ell}_{\rm VN}^{\left(i\right)} = \varnothing$ and ${\bm \ell}_{\rm CN}^{\left(i\right)} = \varnothing$ to record LLR messages with $i=1,2,\cdots,I$;

  Initialize vectors ${\mathbf b}_{\rm VN}^{\left(i\right)}  = \varnothing $ and ${\mathbf b}_{\rm CN}^{\left(i\right)} = \varnothing$ to record code bits, with $i=1,2,\cdots,I$.

  \item \textbf{Count LLR distributions}

  For $k = 1,2,\cdots,K$ and $i=1,2,\cdots,I$, run simulation by using the selected decoding algorithm so as to collect total $E$ LLR messages $\ell_{v \to c}^{\left(i\right)}$ ($\ell_{c \to v}^{\left(i\right)}$) on each edge and their corresponding code bits $x_v$, then append them to ${\bm \ell}_{\rm VN}^{\left(i\right)}$ (${\bm \ell}_{\rm CN}^{\left(i\right)}$), and ${\mathbf b}_{\rm VN}^{\left(i\right)}$ (${\mathbf b}_{\rm CN}^{\left(i\right)}$), respectively;

  For $i=1,2,\cdots,I$, utilize ${\bm \ell}_{\rm VN}^{\left(i\right)}$ (${\bm \ell}_{\rm CN}^{\left(i\right)}$) and ${\mathbf b}_{\rm VN}^{\left(i\right)}$ (${\mathbf b}_{\rm CN}^{\left(i\right)}$) to count PDFs $p_{E,\rm{VN}}\left(\ell\vert 0\right)$ ($p_{E,\rm{CN}}\left(\ell\vert 0\right)$), $p_{E,\rm{VN}}\left(\ell\vert 1\right)$ ($p_{E,\rm{CN}}\left(\ell\vert 1\right)$), and $p_{E,\rm{VN}}\left(\ell\right)$ ($p_{E,\rm{CN}}\left(\ell\right)$) with the histogram method, then calculate the AMI $I_{E,\rm{VN}}^{\left(i\right)}$ ($I_{E,\rm{CN}}^{\left(i\right)}$) according to (\ref{AMI}).

  \item \textbf{Generate the trajectory of T-EXIT}

  On the EXIT plane, write $2I$ AMI pairs as the coordinate points, which can be expressed as:
	\begin{equation}
	\left\{\begin{aligned}
	\left(I_{A,\rm{VN}}^{\left(i\right)},I_{E,\rm{VN}}^{\left(i\right)}\right)=\left(I_{E,\rm{CN}}^{\left(i-1\right)},I_{E,\rm{VN}}^{\left(i\right)}\right),\\
	\left(I_{E,\rm{CN}}^{\left(i\right)},I_{A,\rm{CN}}^{\left(i\right)}\right)=\left(I_{E,\rm{CN}}^{\left(i\right)},I_{E,\rm{VN}}^{\left(i\right)}\right),
	\end{aligned}
	\right.
	\end{equation}
	where $i=1,2,\cdots,I$;
	
	Connect the adjacent coordinate points in the form
    \begin{equation}
      \left(I_{A,\rm{VN}}^{\left(i\right)},I_{E,\rm{VN}}^{\left(i\right)}\right)\to\left(I_{E,\rm{CN}}^{\left(i\right)},I_{A,\rm{CN}}^{\left(i\right)}\right),
    \end{equation}
    where $i=1,2,\cdots,I$. Then, we get the extrinsic AMI transfer trajectory.

    \item \textbf{Generate two transfer functions of T-EXIT}

    Connect the coordinate points in the form
    \begin{equation}
      \left\{ \begin{aligned}
      \left(I_{A,\rm{VN}}^{\left(i\right)},I_{E,\rm{VN}}^{\left(i\right)}\right)\to\left(I_{A,\rm{VN}}^{\left(i+1\right)},I_{E,\rm{VN}}^{\left(i+1\right)}\right),\\
      \left(I_{E,\rm{CN}}^{\left(i\right)},I_{A,\rm{CN}}^{\left(i\right)}\right)\to\left(I_{E,\rm{CN}}^{\left(i+1\right)},I_{A,\rm{CN}}^{\left(i+1\right)}\right),
      \end{aligned}
    \right.
    \end{equation}
    where $i=1,2,\cdots,I-1$. Then, we get two T-EXIT curves with respect to $T_v\left(\cdot\right)$ and $T_c^{-1}\left(\cdot\right)$, respectively.
\end{itemize}

According to the principle of EXIT, there is a transfer relationship between the \emph{a priori} AMI and the extrinsic AMI, i.e.,
\begin{equation}
  I_{A,\rm{CN}}^{\left(i\right)} = I_{E,\rm{VN}}^{\left(i\right)},~ I_{A,\rm{VN}}^{\left(i+1\right)} = I_{E,\rm{CN}}^{\left(i\right)}.
\end{equation}
For each iteration $i$, the extrinsic AMI from VN update $I_{E,\rm{VN}}^{\left(i\right)}$ dominates the system BER performance. According to (\ref{eq_EXIT_functions}), we can also derive that
\begin{equation}\label{eq_EXIT_transfer}
  I_{E,\rm{VN}}^{\left( {i + 1} \right)} = {T_v}\left( {{T_c}\left( {I_{E,\rm{VN}}^{\left( i \right)}} \right)} \right).
\end{equation}
If the iterative decoding finally converges, we have $I_{E,\rm{VN}}^{\left( {i + 1} \right)} = I_{E,\rm{VN}}^{\left( {i} \right)} \triangleq I_{E,\rm{VN}}^{\left( {*} \right)}$. Apparently, $I_{E,\rm{VN}}^{\left( {*} \right)}$ is the \emph{fixed point} of the function in (\ref{eq_EXIT_transfer}), i.e.,
\begin{equation}\label{eq_EXIT_fix_point}
  I_{E,\rm{VN}}^{\left( {*} \right)} = {T_v}\left( {{T_c}\left( {I_{E,\rm{VN}}^{\left( * \right)}} \right)} \right) \Rightarrow T_v^{ - 1}\left( {I_{E,{\rm{VN}}}^{\left( *  \right)}} \right) = {T_c}\left( {I_{E,{\rm{VN}}}^{\left( *  \right)}} \right).
\end{equation}
Followed by this, the coordinate $( {I_{A,{\rm{VN}}}^{\left( *  \right)},I_{E,{\rm{VN}}}^{\left( *  \right)}} )$ of the fixed point on the EXIT plane is the intersection of two curves $T_v\left(\cdot\right)$ and $T_c^{-1}\left(\cdot\right)$. Therefore, after getting these two T-EXIT transfer functions, one can calculate the intersection coordinate that captures the performance at convergence under finite codelength and limited number of iterations.

\section{Performance Evaluation}\label{section_performance_evaluation}

\begin{table*}[htbp]
\renewcommand{\arraystretch}{1.3}
  \centering
  \small
  \caption{BG2 codes for training with the code rate $R = 1/5$.}\label{table_training_length}
  \begin{tabular}{!{\vrule width1pt}m{2.5cm}|m{2cm}|m{2cm}|m{2cm}|m{2cm}!{\vrule width1pt}}
    \Xhline{1pt}
    \centering Index & \centering $C_1$ & \centering $C_2$ & \centering $C_3$ & \centering $C_4$ \tabularnewline
    \hline
    \centering Lifting size & \centering $Z = 3$ & \centering $Z = 6$ & \centering $Z = 10$ & \centering $Z = 16$ \tabularnewline
    \hline
    \centering $\left(N,K\right)$ & \centering $\left(150,30\right)$ & \centering $\left(300,60\right)$ & \centering $\left(500,100\right)$ & \centering $\left(800,160\right)$ \tabularnewline
    \Xhline{1pt}
  \end{tabular}
\end{table*}

\begin{table*}[htbp]
\renewcommand{\arraystretch}{1.3}
  \centering
  \small
  \caption{BG2 codes for training with the number of information bits $K = 160$.}\label{table_training_rate}
  \begin{tabular}{!{\vrule width1pt}m{2.5cm}|m{2cm}|m{2cm}|m{2cm}!{\vrule width1pt}}
    \Xhline{1pt}
    \centering Index & \centering $C_1$ & \centering $C_2$ & \centering $C_3$ \tabularnewline
    \hline
    \centering Code rate $R$ & \centering 0.3701 & \centering 0.3008 & \centering 0.2451 \tabularnewline
    \hline
      \centering $\left(N,K\right)$ & \centering $\left(432,160\right)$ & \centering $\left(532,160\right)$ & \centering $\left(653,160\right)$ \tabularnewline
    \Xhline{1pt}
    \centering Index & \centering $C_4$ & \centering $C_5$ & \centering $C_6$ \tabularnewline
    \hline
    \centering Code rate $R$ & \centering 0.1885 & \centering 0.1533 & \centering 0.1172 \tabularnewline
    \hline
      \centering $\left(N,K\right)$ & \centering $\left(848,160\right)$ & \centering $\left(1043,160\right)$ & \centering $\left(1360,160\right)$ \tabularnewline
    \Xhline{1pt}
  \end{tabular}
\end{table*}

\begin{figure*}[htbp]
\setlength{\abovecaptionskip}{0.cm}
\setlength{\belowcaptionskip}{-0.cm}
\centering{
  \includegraphics[scale=0.72]{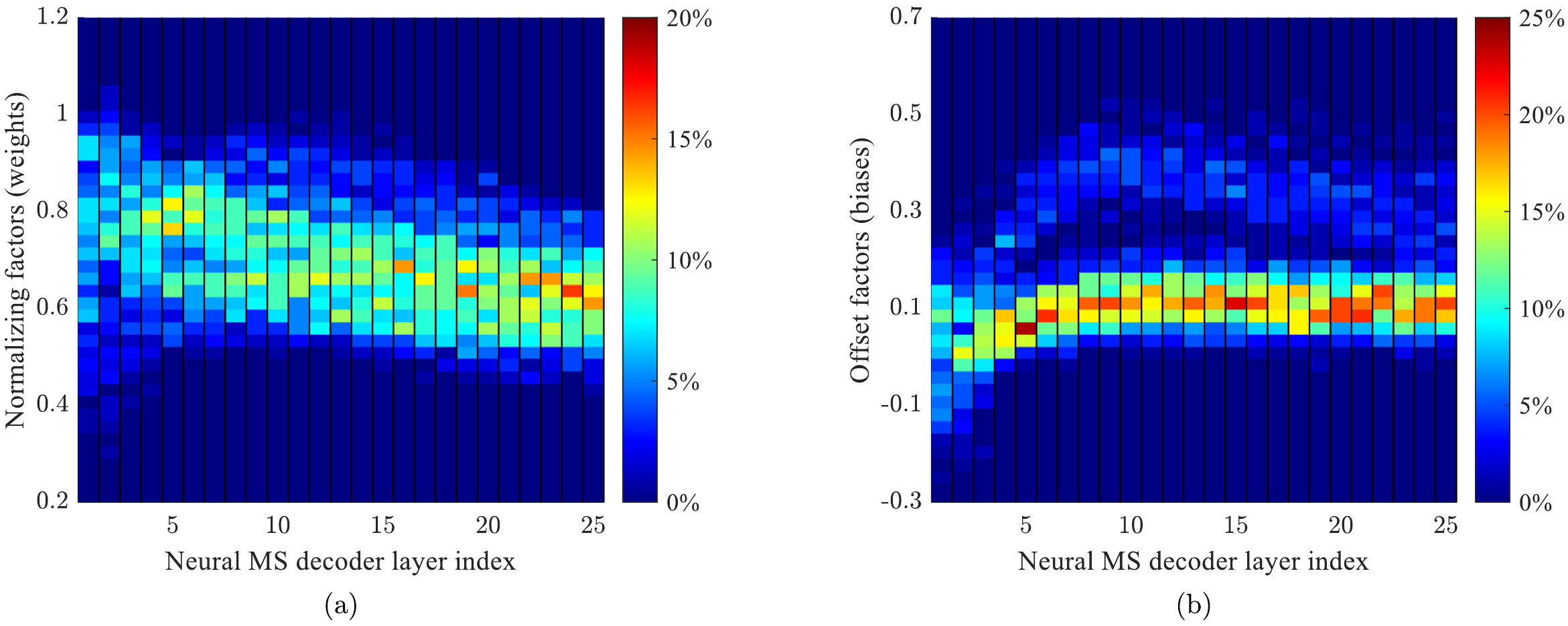}

}
\caption{Distributions of weights and biases for the Type-I neural MS decoder.}\label{fig_distribution_weights_biases_T1}
\end{figure*}

In this section, we construct neural MS decoders for the protograph LDPC codes defined in the 5G standard \cite{5G_NR_std}, and simulate the block error rate (BLER) performance over the binary phase shift keying (BPSK) modulated additive white Gaussian noise (AWGN) channel and the Rayleigh fading channel. The definition of SNR in all these results is the bit SNR, i.e., ${\rm E}_{\rm b}/{N_0}$, where $N_0$ denotes the noise power.

The employed base code is BG2 (defined in \cite[Table 5.3.2-3]{5G_NR_std}), which has 42 rows and 52 columns in the parity-check matrix ${\mathbf H}_{\rm{BG2}}$ with 197 non-zero elements\footnote{One can also select the BG1, and similar results and conclusions can be observed. We do not repeatedly show the performance of codes derived from BG1 in this paper.}. According to the rate-matching algorithm in \cite{5G_NR_std}, the two node-types corresponding to the first two columns in the parity-check matrix ${\mathbf H}_{\rm{BG2}}$ are always punctured, i.e., the code bits of its lifted codes with these two node-types will never be transmitted through the channel. Therefore, the baseline code rate of BG2 codes
is $R = \left( {52 - 42} \right)/\left( {52 - 2} \right) = 1/5$. The lifted codes given in Table \ref{table_training_length} and Table \ref{table_training_rate} are used for training, which are the short and medium codes due to the target codelength of BG2 and limited computational capability. The code rates in Table \ref{table_training_rate} are chosen from the 5G NR modulation and coding scheme (MCS) table \cite[Table 5.1.3.1-1]{5G_NR_std_mcs}. A reference SNR table for training is computed for these codes, which consists of the minimum required SNRs to achieve a target ${\text{BER}}=0.001$ under the AWGN channel and the standard SP decoding with 50 iterations.

The neural MS decoders in Table \ref{table_neural_MS} and Table \ref{table_damp_neural_MS} are trained under the AWGN channel with up to 25 iterations. For each iteration, the parameters are trained over 50000 batches, with 50 samples in each batch. To update the weights/biases/damping factors, we adopt the Adam optimizer \cite{adam} with an initial learning rate of 0.001. The source code used to generate the results is available on github\footnote{The code is available in https://github.com/KyrieTan/Neural-Protograph-LDPC-Decoding}.

\subsection{Training Results}

In this part, we visually demonstrate the trained neural MS decoders. The six types of neural MS decoders in Table \ref{table_neural_MS} and Table \ref{table_damp_neural_MS} are trained by randomly selecting samples from Table \ref{table_training_length}.

For the Type-I neural MS decoder, the distributions of normalizing factors (weights) and offset factors (biases) are presented in Fig. \ref{fig_distribution_weights_biases_T1}(a) and Fig. \ref{fig_distribution_weights_biases_T1}(b), respectively. For each iteration, all the weights or biases are counted as one column in Fig. \ref{fig_distribution_weights_biases_T1}. The weights ${\bm \alpha}^{\left(i\right)}$ vary with iterations, and most values concentrate on the range of 0.5 to 0.75 when $i \ge 9$. For $i < 9$, they concentrate on the range of 0.75 to 0.9. Moreover, as the number of iterations increases, the weights tend to be smaller. The biases ${\bm \beta}^{\left(i\right)}$ focus on the range of 0.05 to 0.15 for any iteration $i \in \left[1,25\right]$.

\begin{figure}[htbp]
\setlength{\abovecaptionskip}{0.cm}
\setlength{\belowcaptionskip}{-0.cm}
  \centering{\includegraphics[scale=0.55]{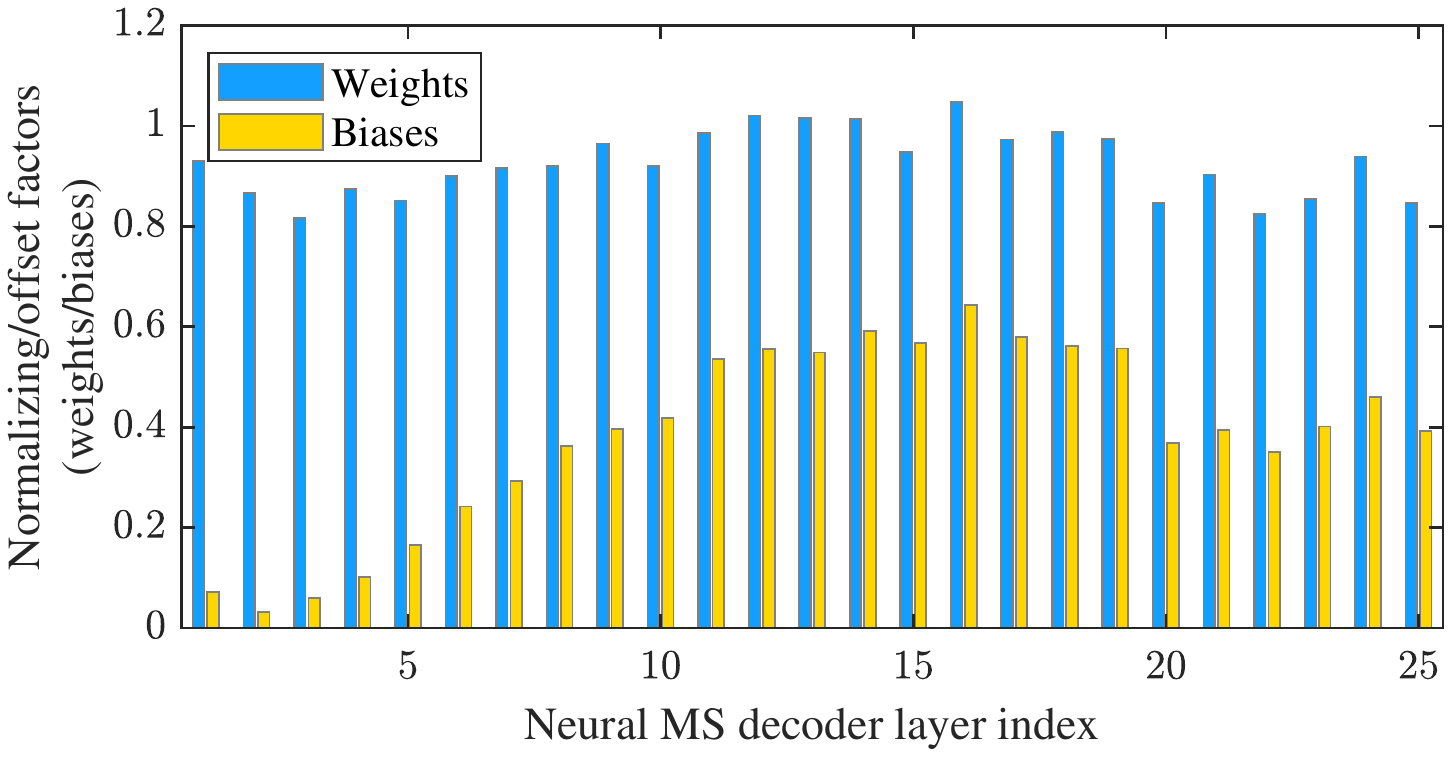}}
  \caption{Plots of weights and biases for the Type-II neural MS decoder.}\label{fig_distribution_weights_biases_T2}
\end{figure}

\begin{figure}[htbp]
\setlength{\abovecaptionskip}{0.cm}
\setlength{\belowcaptionskip}{-0.cm}
  \centering{\includegraphics[scale=0.55]{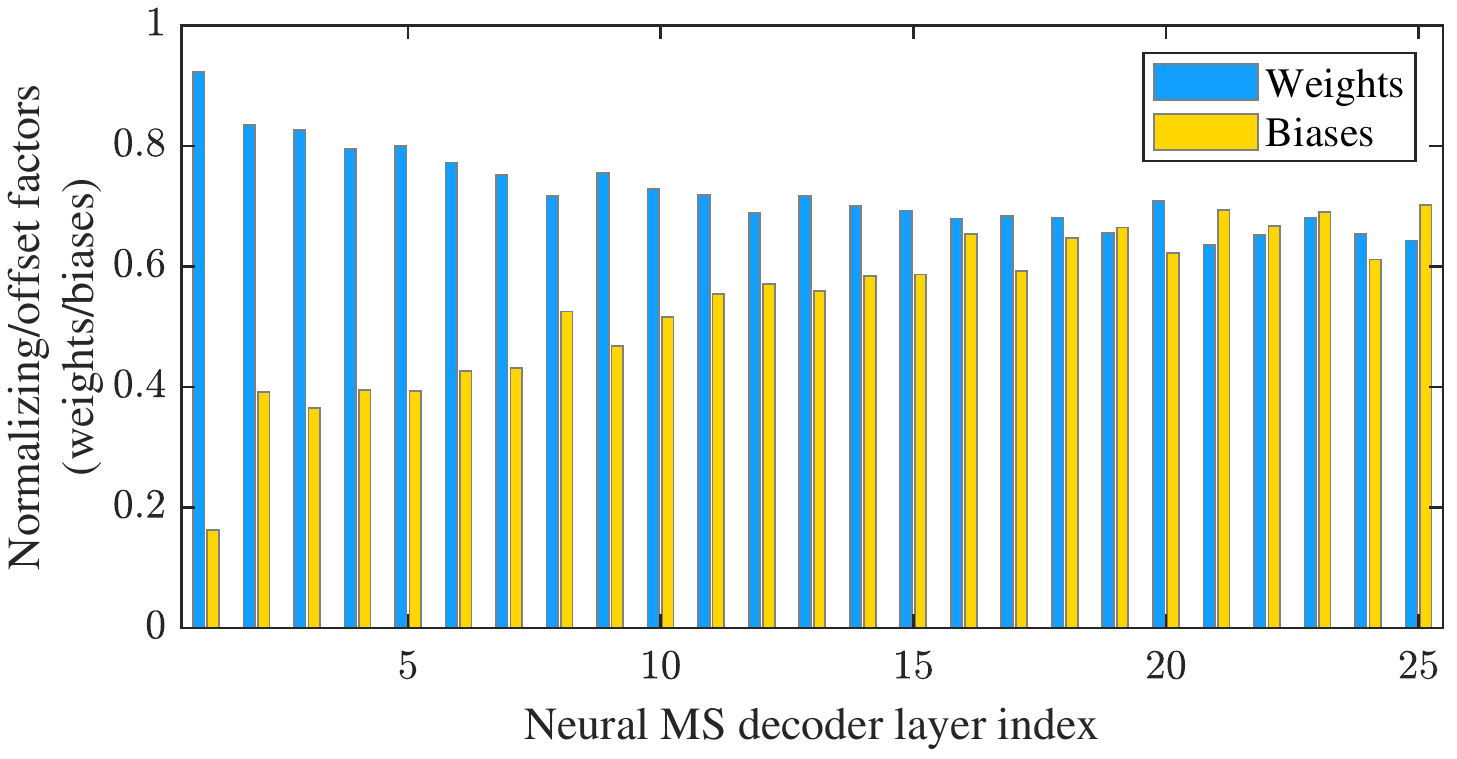}}
  \caption{Plots of weights and biases for the Type-III and Type-IV neural MS decoders, respectively.}\label{fig_distribution_weights_biases_T34}
\end{figure}

For the Type-II neural MS decoder, the plots of weights and biases are shown in Fig. \ref{fig_distribution_weights_biases_T2}. Apparently, compared to the training results of the Type-I decoder, since each iteration only corresponds to one weight ${ \alpha}^{\left(i\right)}$ and bias ${ \beta}^{\left(i\right)}$, we can observe that ${ \alpha}^{\left(i\right)}$ concentrates on the range of 0.8 to 1. However, the bias ${ \beta}^{\left(i\right)}$ varies a lot with the iteration, which increases from 0.05 to 0.6 for $i \le 19$. Then, it stays on a stable value around 0.4 for $i > 19$. That is quiet different from state-of-the-art normalizing and offset factors $\alpha = 0.8$, $\beta = 0.15$ in \cite{minsum_variants} because we jointly optimize these two factors. For the Type-III and Type-IV neural MS decoders, the plots of weights and biases are shown in Fig. \ref{fig_distribution_weights_biases_T34}, respectively. Since only one weight or bias is added to each iteration, that is equivalent to single variable optimization so that the results are different from that in Fig. \ref{fig_distribution_weights_biases_T2}. The weights decrease with iteration $i$, but the biases increase with the iterations.

\begin{figure}[htbp]
\setlength{\abovecaptionskip}{0.cm}
\setlength{\belowcaptionskip}{-0.cm}
  \centering{\includegraphics[scale=0.55]{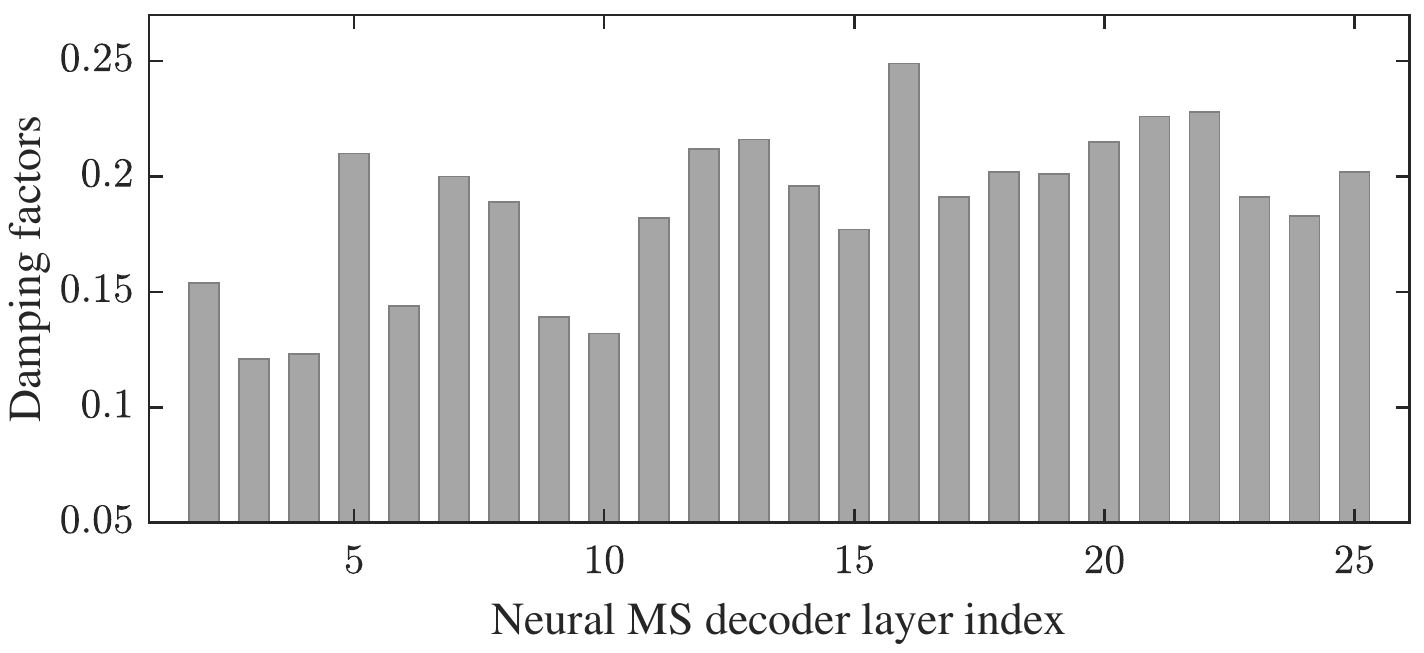}}
  \caption{Plot of damping factors for the Type-VI neural MS decoder.}\label{fig_distribution_damping_T6}
\end{figure}

As for the damping factors of Type-V and Type-VI decoders, we exemplarily show the damping factors of the Type-VI neural MS decoder in Fig. \ref{fig_distribution_damping_T6}. Clearly, all damping factors $\gamma^{\left(i\right)}$ concentrate on the range of 0.15 to 0.2. It means the decoder tends to be less damped, which gives more confidence to the current iteration.

\subsection{T-EXIT Analysis}

\begin{figure*}[htbp]
\setlength{\abovecaptionskip}{0.cm}
\setlength{\belowcaptionskip}{-0.cm}
\centering{
  \includegraphics[scale=0.72]{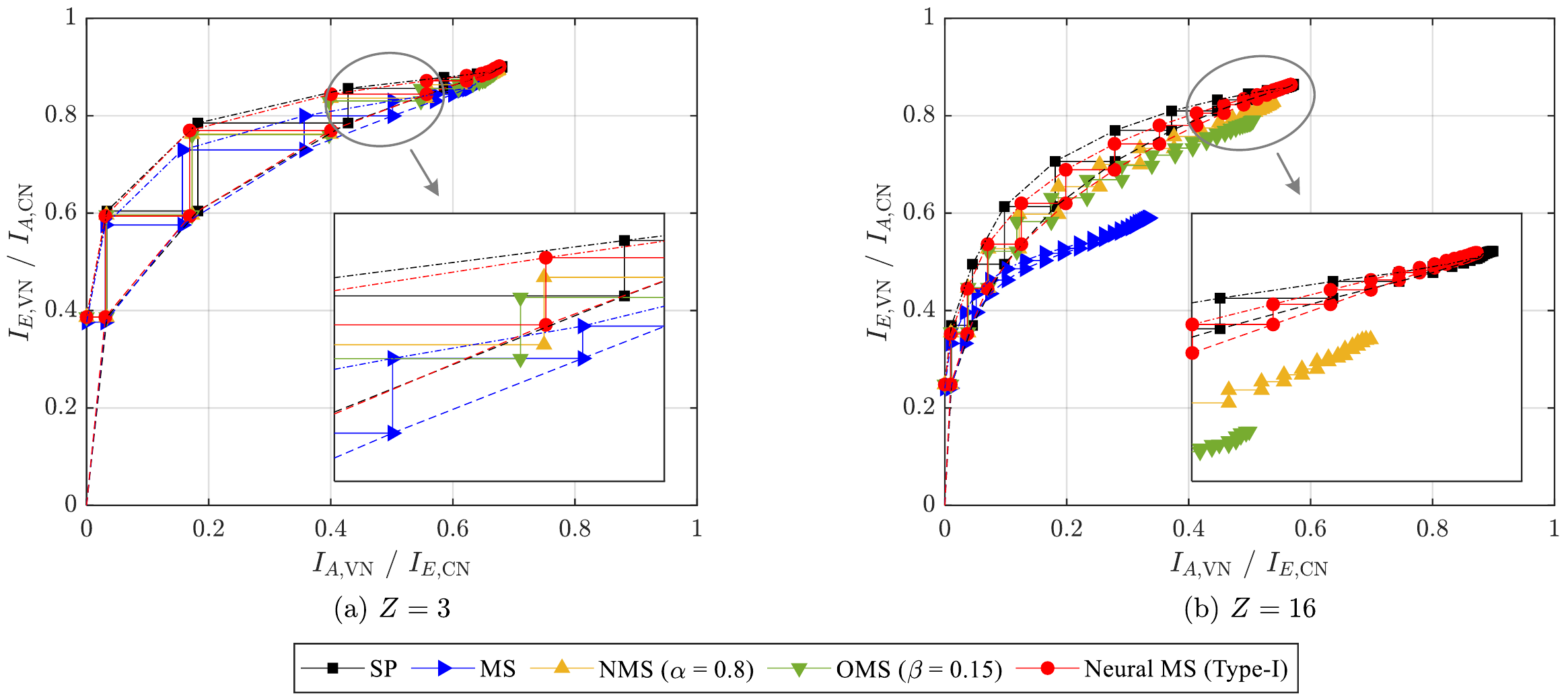}
}
\caption{Two T-EXIT charts for the BG2 LDPC codes under the AWGN channel with $I=25$ iterations. The code with the lifting size $Z=3$ is shown in (a), where the SNR is 4dB. The code with the lifting size $Z=16$ is shown in (b), where the SNR is 1.5dB.}\label{fig_T_EXIT_NN}
\end{figure*}

As a theoretical performance evaluation tool, the T-EXIT analysis results are given in this part. The neural MS decoder is trained by randomly selecting samples from Table \ref{table_training_length}. Two T-EXIT charts for BG2 LDPC codes under the AWGN channel are shown in Fig. \ref{fig_T_EXIT_NN}, where the number of decoding iterations is $I=25$. With $Z=3$ in Fig. \ref{fig_T_EXIT_NN}(a), the extrinsic AMI transfer trajectories of various decoders overlap such that they are somewhat indistinguishable. However, it can be seen from the zoomed area that the extrinsic AMI increment step of the proposed neural MS decoder is superior to others, which only falls behind the SP decoder. When the codelength becomes longer ($Z=16$ in Fig. \ref{fig_T_EXIT_NN}(b)), the difference between extrinsic AMI transfer trajectories becomes more obvious. Although the SP decoder achieves the maximum extrinsic AMI after 25 iterations which corresponds to the best error correction performance, the final extrinsic AMI $I_{E,\rm{CN}}^{\left(25\right)}$ of the neural MS decoder closely follows.

As shown in Table \ref{table_intersection}, we also calculate the intersection coordinate for each pair of T-EXIT curves $T_v\left(\cdot\right)$ and $T_c^{-1}\left(\cdot\right)$ in Fig. \ref{fig_T_EXIT_NN}. The intersection ordinate $I_{E,{\rm{VN}}}^{\left( *  \right)}$ reflects the performance at convergence. For the short codelength with $Z=3$, the proposed neural MS decoding can finally outperform SP decoding and achieve the optimal performance. For the longer codelength with $Z=16$, SP decoding is the most prominent while neural MS decoding closely follows. These results are consistent with the simulation results in Subsection \ref{subsection_simulation}, and imply the superiority of the proposed neural MS decoding.

\begin{table}[htbp]
	\renewcommand{\arraystretch}{1.5}
	\centering
	\small
	\caption{Intersection coordinates of $T_v\left(\cdot\right)$ and $T_c^{-1}\left(\cdot\right)$.}\label{table_intersection}
	\begin{tabular}{!{\vrule width1pt}p{2.4cm}|p{2.3cm}|p{2.3cm}!{\vrule width1pt}}
		\Xhline{1pt}
		\centering \multirow{2}*{\shortstack{Iterative decoding\\algorithm}} & \multicolumn{2}{c!{\vrule width1pt}}{\centering $(I_{A,{\rm{VN}}}^{\left( *  \right)}, I_{E,{\rm{VN}}}^{\left( *  \right)})$} \tabularnewline
		\cline{2-3}
		\centering ~ & \centering $Z = 3$ (4dB) & \centering $Z = 16$ (1.5dB) \tabularnewline
		\Xhline{1pt}
		\centering SP & \centering (0.6806, \bf{0.9013}) & \centering (0.5722, \bf{0.8646}) \tabularnewline
		\hline
		\centering MS & \centering (0.6432, \bf{0.8714}) & \centering (0.3350, \bf{0.5919}) \tabularnewline
		\hline
		\centering NMS & \centering (0.6738, 0.8953) & \centering (0.5388, 0.8302) \tabularnewline
		\hline
		\centering OMS & \centering (0.6698, 0.8928) & \centering (0.5057, 0.7909) \tabularnewline
		\hline
		\centering Neural MS & \centering (0.6764, \bf{0.9014}) & \centering (0.5678, \bf{0.8641}) \tabularnewline
		\Xhline{1pt}
	\end{tabular}
\end{table}

\subsection{Simulation Results}\label{subsection_simulation}

\begin{figure*}[htbp]
\setlength{\abovecaptionskip}{0.cm}
\setlength{\belowcaptionskip}{-0.cm}
\centering{
  \includegraphics[scale=0.72]{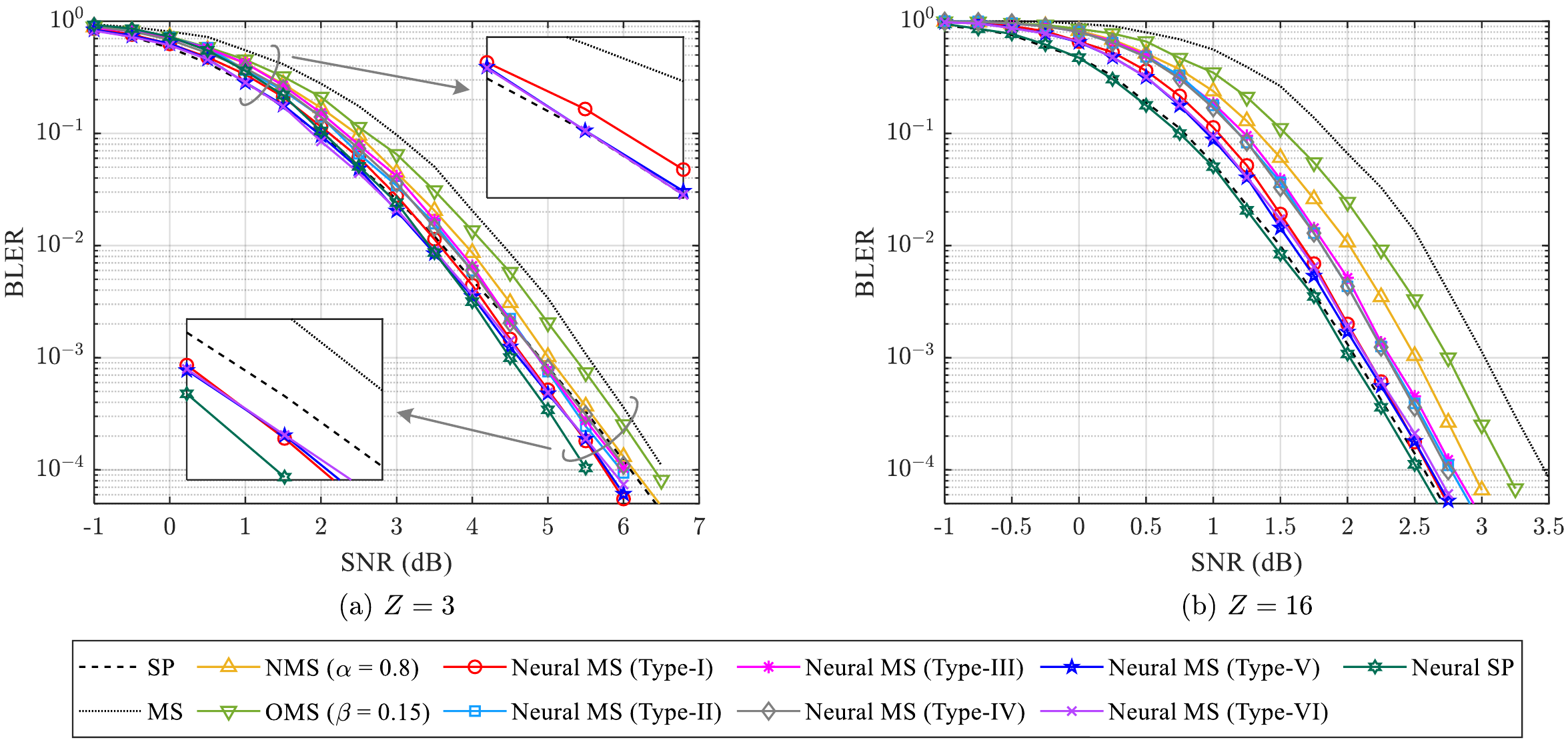}
}
\caption{BLER performance of BG2 codes with lifting sizes $Z=3$ in (a) and $Z=16$ in (b) under the AWGN channel, where the number of iterations is $I=25$.}\label{fig_BLER_AWGN_Z3_16}
\end{figure*}

The BLER results under the AWGN channel with $I=25$ iterations of $\left(150,30\right)$ and $\left(800,160\right)$ BG2 codes are given in Fig. \ref{fig_BLER_AWGN_Z3_16}, whose lifting sizes are $Z=3$ and $Z=16$. The six types of neural MS decoders are trained by randomly selecting samples from Table \ref{table_training_length}. For comparison, the BLER curves of SP, MS, NMS (with a widely-used constant normalizing term $\alpha=0.8$) and OMS (with a widely-used constant offset correction term $\beta=0.15$) decoding algorithms are provided as benchmarks. Note that if we individually optimize $\alpha$ or $\beta$ for each specific code, the performance of NMS and OMS algorithms may be somewhat improved. However, comparing the proposed algorithm with these algorithms is unfair because we apply only one parameter array to multiple codes due to the use of codelength/rate compatible training method. In this figure, we also compare the proposed method with the neural SP decoding method. The two test codes in Fig. \ref{fig_BLER_AWGN_Z3_16} are referred to as the ``matched training and testing'' cases, i.e., they have been selected as samples during the training process.

\begin{figure*}[htbp]
\setlength{\abovecaptionskip}{0.cm}
\setlength{\belowcaptionskip}{-0.cm}
\centering{
  \includegraphics[scale=0.72]{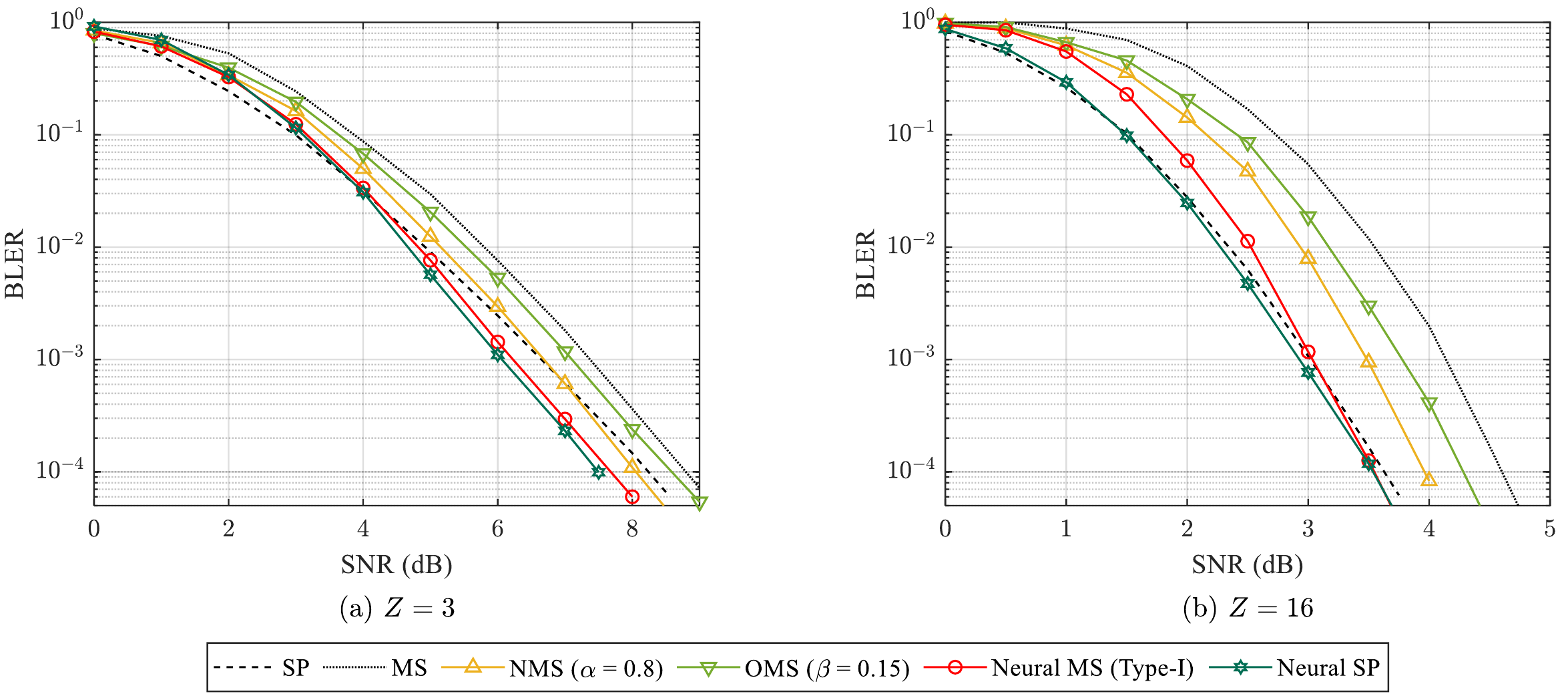}
}
\caption{BLER performance of BG2 codes with lifting sizes $Z=3$ in (a) and $Z=16$ in (b) under the Rayleigh fading channel, where the number of iterations is $I=25$.}\label{fig_BLER_Fading_Z3_16}
\end{figure*}

For the short code of $Z=3$, we observe that all neural MS decoders present gains with respect to the original MS decoder. The Type-I neural MS decoder outperforms OMS and NMS by 0.4dB and 0.2dB, respectively. Equipped with damping factors, the performance of Type-V\&VI neural MS decoders can be further improved in the low SNR region. More interestingly, it can even outperform SP decoding in the high SNR region. The reasons for this performance gain are twofold as explained in Subsection \ref{subsection_gains_explanation}. As for the longer code with $Z=16$, the gains of neural MS decoders versus the original MS decoder become larger, and the Type-I neural MS decoder outperforms OMS and NMS by 0.5dB and 0.3dB, respectively. Compared to SP decoding, although there exists some slight performance loss of the Type-I neural MS decoder, it involves much lower complexity and quite closely approaches SP decoding in the high SNR region. Meanwhile, the trained decoders show good generalization ability to different codelengths due to the length compatible training method. In addition, from the results of damped neural MS decoders, we find that using per-edge-type damping factors (Type-V) does not improve much upon using single parameter per-iteration (Type-VI). Hence, the Type-VI neural MS decoder with only one damping factor $\gamma^{\left(i\right)}$ on each iteration achieves a better tradeoff between performance and complexity.

Fig. \ref{fig_BLER_Fading_Z3_16} shows the BLER results under the Rayleigh fading channel, where the code configurations are the same as that in Fig. \ref{fig_BLER_AWGN_Z3_16}. The neural network parameters trained under the AWGN channel in Fig. \ref{fig_BLER_AWGN_Z3_16} are directly used in this figure. It is interesting to observe that the proposed neural MS decoding still performs well especially in the high SNR region. With $Z = 16$, the proposed neural MS decoder outperforms the standard SP decoding, which is different from that under the AWGN channel. These results imply that the trained neural decoder presents good robustness under the Rayleigh fading channel.

\begin{figure*}[htbp]
\setlength{\abovecaptionskip}{0.cm}
\setlength{\belowcaptionskip}{-0.cm}
\centering{
  \includegraphics[scale=0.72]{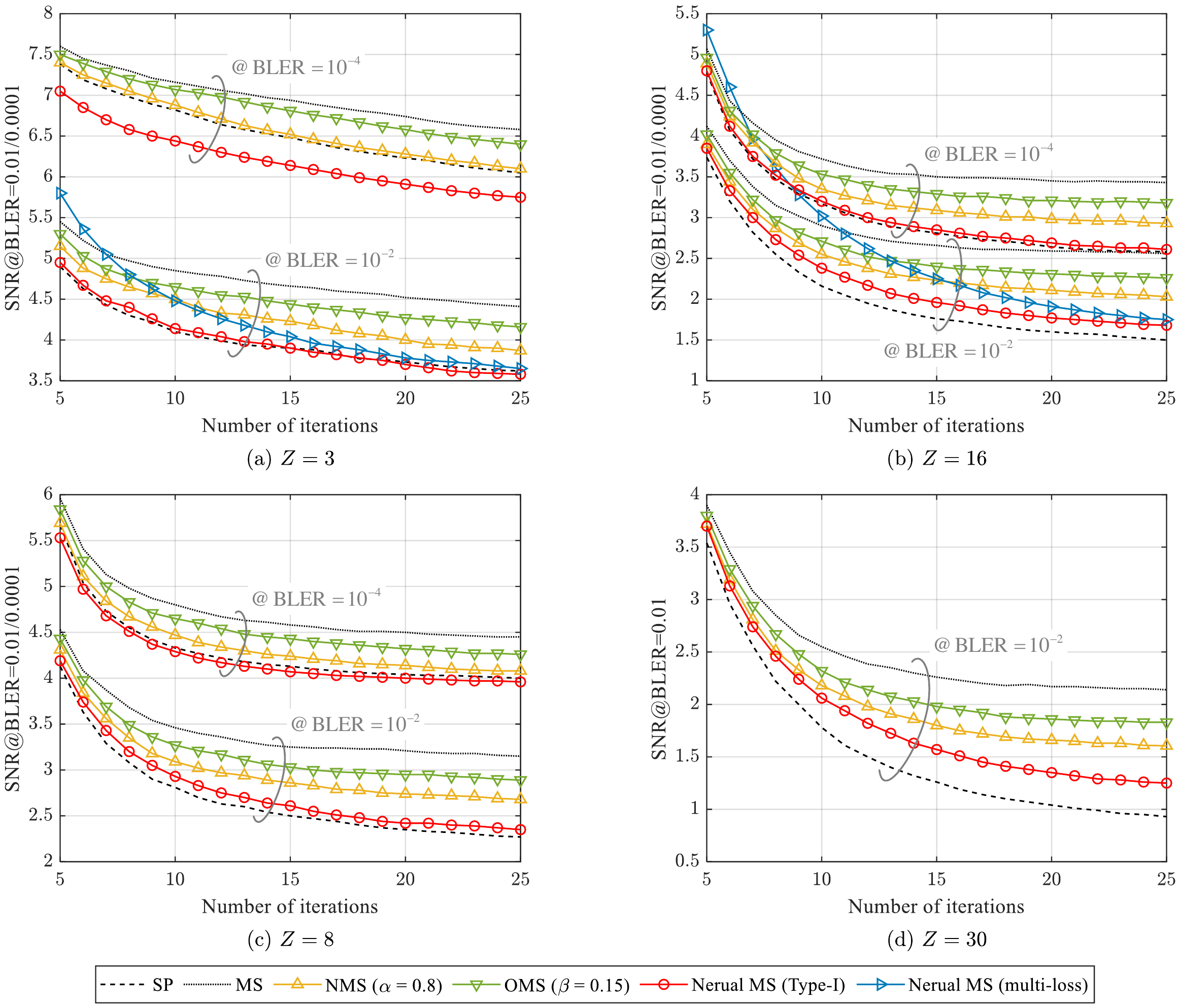}
}
\caption{SNR required to achieve ${\text{BLER}} = {10^{-2}}~{\text{or}}~{10^{-4}}$ under the AWGN channel, (a) shows the $\left(150,30\right)$ code with $Z=3$, (b) shows the $\left(800,160\right)$ code with $Z=16$, (c) shows the $\left(400,80\right)$ code with $Z=8$, and (d) shows the $\left(1500,300\right)$ code with $Z=30$.}\label{fig_SNR_AWGN_Z3_16_8_30}
\end{figure*}

Fig. \ref{fig_SNR_AWGN_Z3_16_8_30} shows the SNR required for various decoders with a certain number of iterations to achieve ${\text{BLER}} = {10^{-2}}~{\text{or}}~{10^{-4}}$ under the AWGN channel. The neural MS decoder is trained by randomly selecting samples from Table \ref{table_training_length}. The results in Fig. \ref{fig_SNR_AWGN_Z3_16_8_30}(a) ($Z = 3$) and Fig. \ref{fig_SNR_AWGN_Z3_16_8_30}(b) ($Z = 16$) correspond to the ``matched training and testing'' cases, and the results in Fig. \ref{fig_SNR_AWGN_Z3_16_8_30}(c) ($Z = 8$) and Fig. \ref{fig_SNR_AWGN_Z3_16_8_30}(d) ($Z = 30$) correspond to the ``mismatched training and testing'' cases, i.e., these two lifting sizes are not included in the training set of Table \ref{table_training_length}.

\begin{figure*}[htbp]
\setlength{\abovecaptionskip}{0.cm}
\setlength{\belowcaptionskip}{-0.cm}
\centering{
  \includegraphics[scale=0.72]{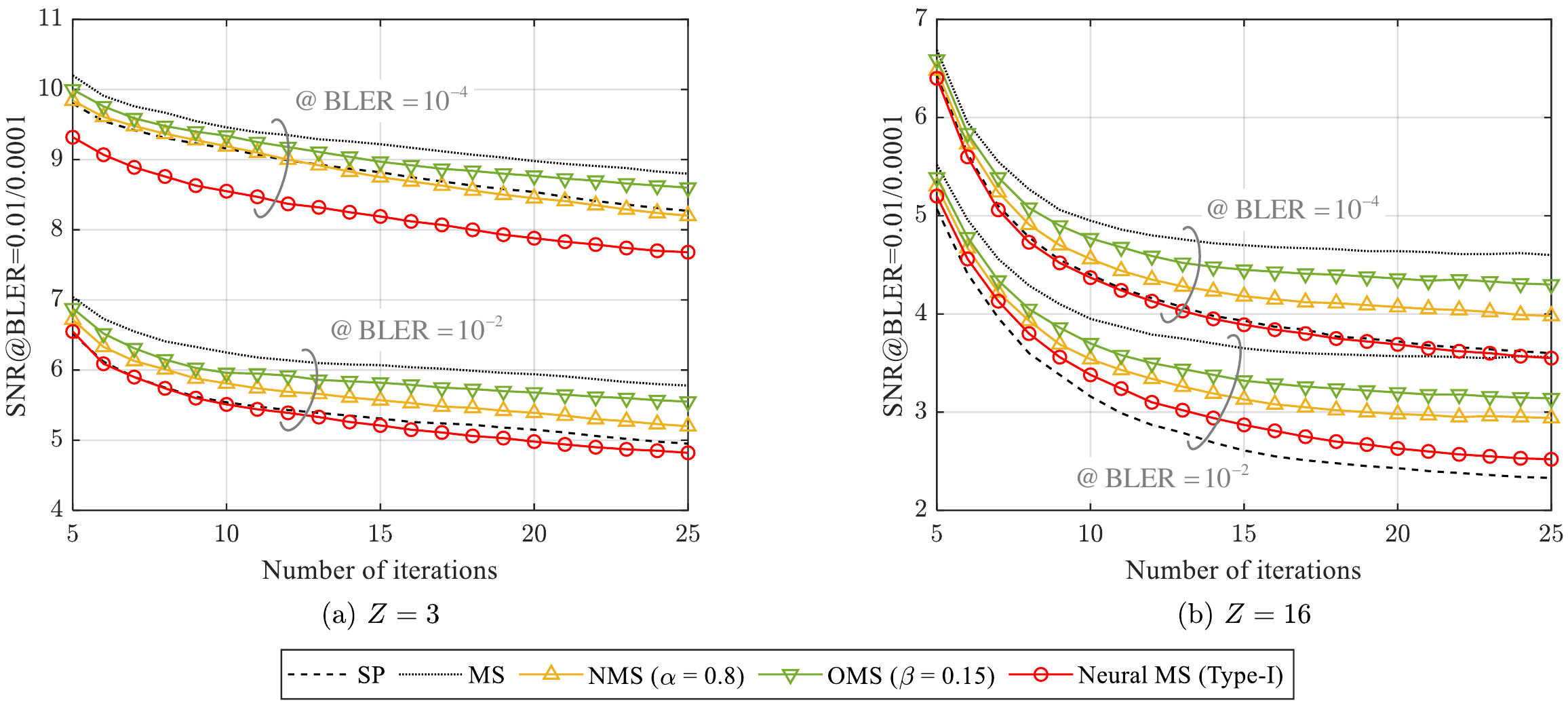}
}
\caption{SNR required to achieve ${\text{BLER}} = {10^{-2}}~{\text{or}}~{10^{-4}}$ under the Rayleigh fading channel, (a) shows the $\left(150,30\right)$ code with $Z=3$, (b) shows the $\left(800,160\right)$ code with $Z=16$ in (b).}\label{fig_SNR_Fading_Z3_16}
\end{figure*}

\begin{figure*}[htbp]
\setlength{\abovecaptionskip}{0.cm}
\setlength{\belowcaptionskip}{-0.cm}
\centering{
  \includegraphics[scale=0.72]{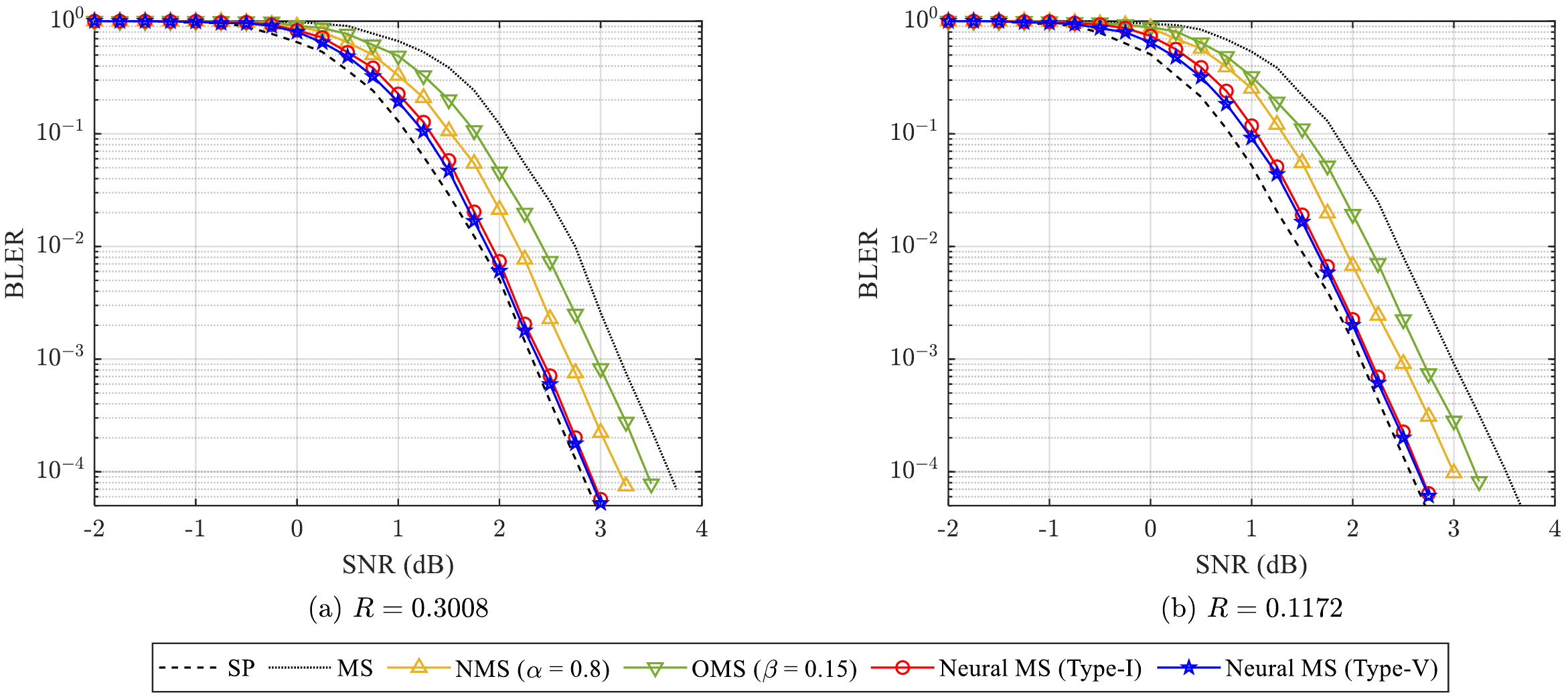}
}
\caption{BLER performance of BG2 codes under the AWGN channel with $K=160$ and code rates $R=0.3008$ (i.e., $\left(532,160\right)$ code) in (a) and $R=0.1172$ (i.e., $\left(1360,160\right)$ code) in (b), the number of iterations $I=25$.}\label{fig_BLER_AWGN_Diff_Rate_Z16}
\end{figure*}

With $Z=3$ in Fig. \ref{fig_SNR_AWGN_Z3_16_8_30}(a), considering the SNR required to achieve ${\text{BLER}} = {10^{-2}}$, the neural MS decoder outperforms the standard MS decoder more than 0.5dB, and it is quite close to the standard SP decoder. Fig. \ref{fig_SNR_AWGN_Z3_16_8_30}(a) also gives the required SNRs to achieve ${\text{BLER}} = {10^{-4}}$, in this high SNR region, the neural MS decoder can even outperform the SP decoder at early iterations. In this case, we also show the performance of the neural MS decoder under the conventional multi-loss training \cite{deep_learning_decoding_JSTSP} with $I = 25$. The proposed greedy training method tends to optimize the performance of neural decoders as early as possible, i.e., faster convergence. However, the traditional multi-loss training \cite{deep_learning_decoding_JSTSP} tends to optimize the performance for the target number of iterations, i.e., good results can only be observed at the final output. Hence, given $I = 25$, we can observe that these two training methods finally achieve similar performance when the number of iterations finally reaches 25, though the proposed training method still presents some gain at this point. Fig. \ref{fig_SNR_AWGN_Z3_16_8_30}(b) shows the required SNRs to achieve ${\text{BLER}} = {10^{-2}}$ when $Z=16$. Though still better than the MS/NMS/OMS algorithms, the performance of neural MS decoding is slightly inferior to SP decoding. For the required SNRs to achieve ${\text{BLER}} = {10^{-4}}$ in Fig. \ref{fig_SNR_AWGN_Z3_16_8_30}(b), we observe that the neural MS decoder and the SP decoder achieve the same performance, which implies the superiority of the neural MS deocder in the high SNR region.

The results of mismatched training and testing are shown in Fig. \ref{fig_SNR_AWGN_Z3_16_8_30}(c) and Fig. \ref{fig_SNR_AWGN_Z3_16_8_30}(d). The proposed neural MS decoder still provides stable performance gain versus MS decoding and approach SP decoding, which implies the generalization ability of the neural MS decoder. With $Z = 8$, the proposed neural MS decoding method outperforms SP decoding at ${\text{BLER}} = {10^{-4}}$. For longer codes, e.g., $Z =30$, it may not outperform SP decoding. The explanations about these observations are given in Subsection \ref{subsection_gains_explanation}.

Fig. \ref{fig_SNR_Fading_Z3_16} shows the SNR required for various decoders to achieve ${\text{BLER}} = {10^{-2}}~{\text{or}}~{10^{-4}}$ under the Rayleigh fading channel. From this figure, we can see that, with $Z = 3$, the performance gain of neural MS decoding versus SP decoding can reach 0.6dB at ${\text{BLER}} = {10^{-4}}$, which is larger than that of 0.4dB under the AWGN channel. With $Z = 16$, neural MS decoding also outperforms SP decoding at ${\text{BLER}} = {10^{-4}}$.

The BLER results with $I=25$ iterations of $\left(532,160\right)$ and $\left(1360,160\right)$ BG2 codes are given in Fig. \ref{fig_BLER_AWGN_Diff_Rate_Z16}, whose code rates are $R=0.3008$ and $R=0.1172$, respectively. All these neural MS decoders are trained by randomly selecting samples from Table \ref{table_training_rate}, and these code rates are chosen from the 5G NR MCS table \cite[Table 5.1.3.1-1]{5G_NR_std_mcs}. We also observe that the neural MS decoder outperforms the traditional MS/NMS/OMS decoders. In addition, it approaches SP decoding in the high SNR region. These results imply the generalization ability of the neural MS decoder to different code rates due to the rate compatible training method.

\subsection{Discussion about the Gains of Neural MS Decoding}\label{subsection_gains_explanation}

Combing all the simulation results, we can summarize the following conclusions:
\begin{itemize}
  \item For short codelengths, e.g., $Z=3$, the proposed neural MS decoding method can provide superior BLER performance with respect to the standard MS/NMS/OMS decoding methods. In the high SNR region, it can even outperform the SP decoding method.

  \item For medium or long codelengths, e.g., $Z=16$, the proposed neural MS decoding can approach the performance of SP decoding but may not surpass SP.
\end{itemize}

To explain the gains achieved by the proposed neural MS decoding, in Table \ref{table_cycles}, we count the number of short cycles of the aforementioned codes. Clearly, some short cycles exist, e.g., 4-cycles, especially for short codes. As the lifting size increases, the number of 4-cycles decreases rapidly.

\begin{table}[htbp]
\renewcommand{\arraystretch}{1.3}
  \centering
  \small
  \caption{Number of short cycles for different lifting sizes $Z$ of BG2.}\label{table_cycles}
  \begin{tabular}{!{\vrule width1pt}m{1cm}|m{1.3cm}|m{1.3cm}|m{1.3cm}|m{1.3cm}!{\vrule width1pt}}
    \Xhline{1pt}
    \centering \multirow{2}*{\shortstack{Cycle\\length}} & \multicolumn{4}{c!{\vrule width1pt}}{Lifting size} \tabularnewline
    \cline{2-5}
    \centering ~ & \centering $Z = 3$ & \centering $Z = 8$ & \centering $Z = 16$ & \centering $Z = 30$ \tabularnewline
    \Xhline{1pt}
    \centering 4 & \centering 428 & \centering 224 & \centering 176 & \centering 0 \tabularnewline
    \hline
    \centering 6 & \centering 11511 & \centering 11800 & \centering 10768 & \centering 11460 \tabularnewline
    \hline
    \centering 8 & \centering 339849 & \centering 373044 & \centering 379192 & \centering 372750 \tabularnewline
    \hline
    \centering 10 & \centering 9823374 & \centering 11642984 & \centering 11926672 & \centering 12082290 \tabularnewline
    \Xhline{1pt}
  \end{tabular}
\end{table}

\begin{table*}[htbp]
\renewcommand{\arraystretch}{1.3}
  \centering
  \small
  \caption{Operations required for one iteration.}\label{table_complexity}
  \begin{tabular}{!{\vrule width1pt}m{3.5cm}!{\vrule width1pt}m{1.5cm}|m{1.5cm}|m{1.5cm}|m{1.5cm}|m{1.5cm}!{\vrule width1pt}}
    \Xhline{1pt}
    \centering Operation & \centering $\tanh$ & \centering $\times$ & \centering $+$ & \centering comp. & \centering sign flip \tabularnewline
    \Xhline{1pt}
    \centering SP & \centering $2E$ & \centering $\approx 2E$ & \centering $\approx 2E$ & \centering -- & \centering -- \tabularnewline
    \hline
    \centering MS & \centering -- & \centering -- & \centering $\approx 2E$ & \centering $\approx 2E$ & \centering $\approx 2E$ \tabularnewline
    \hline
    \centering NMS & \centering -- & \centering $E$ & \centering $\approx 2E$ & \centering $\approx 2E$ & \centering $\approx 2E$ \tabularnewline
    \hline
    \centering OMS & \centering -- & \centering -- & \centering $\approx 3E$ & \centering $\approx 2E$ & \centering $\approx 2E$ \tabularnewline
    \hline
     \centering Type-I\&II neural MS & \centering -- & \centering $E$ & \centering $\approx 3E$ & \centering $\approx 2E$ & \centering $\approx 2E$ \tabularnewline
    \hline
    \centering Type-III neural MS & \centering -- & \centering $E$ & \centering $\approx 2E$ & \centering $\approx 2E$ & \centering $\approx 2E$ \tabularnewline
    \hline
    \centering Type-IV neural MS & \centering -- & \centering -- & \centering $\approx 3E$ & \centering $\approx 2E$ & \centering $\approx 2E$ \tabularnewline
    \hline
     \centering Type-V\&VI neural MS & \centering -- & \centering $2E$ & \centering $\approx 4E$ & \centering $\approx 2E$ & \centering $\approx 2E$ \tabularnewline
     \hline
     \centering Neural SP & \centering $2E$ & \centering $\approx 3E$ & \centering $\approx 3E$ & \centering -- & \centering -- \tabularnewline
    \Xhline{1pt}
  \end{tabular}
\end{table*}

Following previous works \cite{minsum_variants} and \cite{deep_learning_decoding_JSTSP}, the gains achieved by the proposed algorithm stems from two aspects:

\subsubsection{Better Approximation to the SP Algorithm}

Well-learned weights (normalizing factors) and biases (offset factors) make the neural MS decoding method a better approximation to the SP algorithm. This property contributes to most of the performance gain. It is inherently aligned with the conventional optimization of MS decoding by introducing one normalizing or offset factor \cite{minsum_variants}, which should be individually optimized by using the DE algorithm for each specific LDPC code \cite{minsum_DE}. In our work, we can optimize multiple weights and biases in parallel, and thus the degrees of freedom for optimization are expanded. Therefore, more gains can be observed with respect to the single parameter optimization.

\subsubsection{Mitigating the Effects of Short Cycles}

Well-learned parameters can further mitigate the effects of short cycles, which is aligned with the performance gain stated by Nachmani \emph{et al.} in \cite{deep_learning_decoding_JSTSP}. Indeed, this can be seen more clearly for HDPC codes as seen in \cite{deep_learning_decoding_JSTSP}. This inference is based on the fact that there exist a large number of short cycles in the Tanner graph of HDPC codes. In comparison, sophisticated LDPC codes may efficiently eliminate short cycles; nevertheless, in this paper, we design neural MS decoding for protograph LDPC codes. For this special but widely-used case, each code is lifted from the base graph, which adds flexibility while also inevitably introducing some short cycles, e.g., 4-cycles. From Table \ref{table_cycles}, we find that there indeed exist some short cycles, especially for short codes, i.e., small lifting sizes. From this conceptual perspective, we reasonably interpret the additional gain achieved by the proposed neural MS decoding.

For medium or long codelengths, as shown in Table \ref{table_cycles}, the number of short cycles decreases a lot so that SP decoding presents good performance. At this time, the main task of the neural MS decoder is to compensate for the approximation loss of MS decoding so that it may not surpass SP decoding.

We present the neural SP decoding results as a supplementary proof of the above explanations in Fig. \ref{fig_BLER_AWGN_Z3_16}. The results show that it can also beat standard SP decoding. Combining with the cycle distribution results in Table \ref{table_cycles}, we verify the statement of restraining short cycles by neural decoding. However, we also observe that the performance gain of neural SP decoding with respect to the proposed neural MS decoding is trivial in the high SNR region, i.e., the BLER performance of neural MS decoding is quite close to that of neural SP decoding. It means that well-learned weights and biases in the neural MS decoder are already enough to compensate for the loss of MS approximation and mitigate the effects of short cycles like that in the neural SP decoder.

\subsection{Complexity Comparison}

Table \ref{table_complexity} summarizes the required operations for one iteration of the six types of neural MS decoder in Table \ref{table_neural_MS} and Table \ref{table_damp_neural_MS}, along with SP, MS, NMS, and OMS algorithms. $E$ denotes the number of edges in the Tanner graph of LDPC codes. The neural MS decoders avoid computing hyperbolic tangent functions, and they have a complexity roughly the same with NMS and OMS. Since different weights and biases are engaged per iteration, the neural MS decoders require additional memory to store the parameters. Also, it should be noted that the cost of damping operation in the Type-V\&VI decoders is not only the additional $2E$ multiplications ``$\times$'' but also the additional memory to store the previous iteration's LLRs. The neural SP decoder is of the highest complexity.

\section{Conclusion and Future Trends}\label{section_conclusion}

In this paper, a set of neural MS decoding algorithms for protograph LDPC codes has been presented. In these algorithms, weights and biases are added to the edges corresponding to the CN to VN updating equation of the MS algorithm, which renders neural MS decoding a better approximation to the SP. To mitigate the impact of short cycles in iterative decoding, the weights and biases are allowed to take different values for different iterations. Exploiting the lifting structure of protograph LDPC codes, the same parameter has been shared among a bundle of edges which are derived from the same edge in the base graph. A low-complexity iteration-by-iteration training mechanism has been proposed for tuning the parameters, which also avoids the vanishing gradient problem. The proposed training method enables the neural MS decoder to achieve better performance with faster convergence. Combined with damping factors, the performance of the neural MS decoder can be further improved. The proposed neural MS decoders have similar complexity compared to MS/NMS/OMS and are much more hardware-friendly than SP. Results have shown that the performance of neural MS decoding is consistently better than original MS/NMS/OMS algorithms, and is close to or even better than the SP algorithm.

There exist many issues to be addressed in the future. One of them is the quantization of neural decoder parameters, we can investigate the influence of quantized parameters on the neural MS decoder. In addition, the similar method to compensate for the loss of message quantization in hardware implementation of protograph LDPC decoding is worth to be studied.

\section*{Acknowledgment}

The authors would like to thank Kai Chen and Dexin Zhang for their suggestions on this work.

\ifCLASSOPTIONcaptionsoff
  \newpage
\fi

\end{document}